\documentclass[10pt,preprint,authoryear,longnamesfirst]{aastex}
\usepackage{psfig}
\received{2000 September~8}
\revised{2001 April~26}
\accepted{2001 May~17}
\paperid{52620}
\cpright{PD}{2000}
\shorttitle{Compact Radio Source Variability}
\shortauthors{Lazio et al.}

\begin{document}

\title{A Dual Frequency, Multi-Year Monitoring Program of Compact
	Radio Sources}
\author{T.~Joseph~W.~Lazio, E.~B.~Waltman}
\affil{Code~7213, Remote Sensing Division, Naval Research Laboratory, 
	 Washington, DC 20375-5351}
\email{lazio@rsd.nrl.navy.mil}
\email{ewaltman@rsd.nrl.navy.mil}

\author{F.~D.~Ghigo}
\affil{National Radio Astronomy Observatory, 
	\hbox{P.{}O.}~Box~2, Green Bank, WV  24944}
\email{fghigo@nrao.edu}

\author{R.~L.~Fiedler}
\affil{Code~7260, Remote Sensing Division, Naval Research Laboratory,
	 Washington, DC 20375-5351}
\email{fiedler@sealab.nrl.navy.mil}

\author{R.~S.~Foster\altaffilmark{1}}
\affil{Code~7213, Remote Sensing Division, Naval Research Laboratory,
	 Washington, DC 20375-5351}

\and 

\author{K.~J.~Johnston}
\affil{US Naval Observatory, 3450 Massachusetts Ave.~NW,
	 Washington, DC 20392-5420}
\email{ken@spica.usno.navy.mil}

\altaffiltext{1}{Present address: Roger S.~Foster, Booz-Allen \&
Hamilton Inc., 8283~Greensboro Drive, McLean, VA  22102-3838 USA}

\begin{abstract}
We present light curves for 149~sources monitored with the Green Bank
Interferometer.  The light curves are at two radio frequencies
(approximately 2.5 and~8.2~GHz) and range from~3 to~15~yrs in length,
covering the interval 1979--1996, and have a typical sampling of one
flux density measurement every 2~days.  We have used these light
curves to conduct various variability analysis (rms flux density
variations and autoregressive, integrated, moving average modeling) of
these sources.  We find suggestive, though not unambiguous evidence,
that these sources have a common, broadband mechanism for intrinsic
variations, in agreement with previous studies of a subset of these
source.  We also find that the sources generally display a short-term
variability ($\sim 10$~d) that arises from radio-wave scattering in an
extended medium.  These conclusions extend those of
\citeauthor*{fiedleretal87} who used a sub-sample of these data.  The
primary motivation for this monitoring program was the identification
of extreme scattering events.  In an effort to identify ESEs in a
systematic manner, we have taken the wavelet transform of the light
curves.  We find 15 events in the light curves of~12 sources that we
classify as probable ESEs.  However, we also find that five ESEs
previously identified from these data do not survive our wavelet
selection criteria.  Future identification of ESEs will probably
continue to rely on both visual and systematic methods.  Instructions
for obtaining the data are also presented.
\end{abstract}


\section{Introduction}\label{sec:intro}

Radio light curves of extragalactic sources can be a powerful probe of
both intrinsic and propagation-induced variability of the sources.
The typical variability time scale of both intrinsic activity and
propagation effects requires that multi-year monitoring programs be
conducted.  A variety of factors make such multi-year monitoring
programs difficult to sustain, but a number have been conducted.
Recent examples include \cite{gg-k90,haa92,vtunlv92,bpgmss94} and
\cite{sagdmp99} (see also \S1 of \citealt*{fiedleretal87}).

With respect to intrinsic activity, previous monitoring programs have
suggested, though not entirely unambiguously, that the emission from
different classes of sources may be produced by the same mechanism, as
would be expected from various unification scenarios.  Reasonable
agreement is found in comparisons of the predicted light curves from
shocked jet models \citep{mg85,haa89a} and observed light curves of
various sources (\objectname[]{BL~Lac}, \citealt*{haa89b};
\objectname[3C]{3C~279} and \objectname[Ohio]{OT~081},
\citealt{haa91}).  In these models, the emission results from a
synchrotron-emitting plasma propagating along a diverging jet and
subject to one or more shocks.  \citet*{fiedleretal87} performed a
correlation analysis on light curves of a sample of 36 sources and
found that BL~Lacs and radio-loud quasars had similar correlation
functions, suggesting that similar processes produced the emission in
both classes of sources.  A structure-function analysis of the total
intensity of 51~sources monitored by the University of Michigan Radio
Astronomy Observatory \citep{haa92} also showed that BL~Lac objects
and radio-loud quasars exhibit similar, power-law structure functions.

However, these comparisons, particularly those of \cite{haa89a} and
\cite{haa91}, indicated that there are differences in the radio
emission process between radio-loud quasars and BL~Lacs.  \cite{haa92}
also found that a structure-function analysis of the Stokes parameters
in the frame of reference of the VLBI jet indicated that variations of
the polarized flux occur in an orientation related to the underlying
jet direction, though with a magnetic field less strongly correlated
with jet direction for radio-loud quasars than for BL~Lac objects.
Flux-based selection effects, due to opacity and Doppler boosting,
tend to discriminate against observing the strongest polarization
assumed to be present in these categories of sources.  Finally,
\citet{aah92} concluded that the statistical polarization of 62
Pearson-Readhead sources in the UMRAO monitoring program indicated
intrinsic differences between BL~Lac objects and radio-loud quasars.

With respect to propagation-induced variability, multi-year monitoring
programs of compact sources have shown month- to year-long variations
thought to be due to refractive focussing and defocussing of the
propagating radiation \citep{rcb84}.  The refractive variations are
caused by plasma density fluctuations having typical scale sizes of
order an astronomical unit.  Propagation-induced variability in a
monitoring program can provide information on the angular and/or
spatial distribution of these density fluctuations and potentially
constrain the microphysics responsible for generating or maintaining
these structures.  In one case, \cite{bpgmss94} found an annual
variation in the scattering properties, which they attributed to
changes in the line of sight to sources during the course of the
Earth's orbit.

A more extreme example of propagation-induced variability is extreme
scattering events (ESE)---a class of dramatic ($\gtrsim 50$\%)
decreases in the flux density of extragalactic sources near 1~GHz
which persist for weeks to months and are bracketed by substantial
increases \citep{fdjh87}.  ESEs are infrequent, occurring roughly 1\%
of the time \citep{fdjws94}, so only a multi-year program observing a
large number of sources will have a reasonable probability of
detecting one.

This paper reports light curves for 149 sources monitored during the
interval 1988--1994.4, primarily to search for ESEs.  This sample
expands upon and includes the 36 sources monitored by
\citeauthor*{fiedleretal87}~(\citeyear{fiedleretal87}, hereinafter
\citeauthor{fiedleretal87}) during the interval 1979--1985.  
\citeauthor{fiedleretal87} presented light curves, correlation
functions, and structure functions for that 36-source subsample.
Light curves for a 46-source subsample monitored during the interval
1979--1987 were presented by \citet{wfjsfjmm91}.  In addition,
interstellar scintillation studies and ESE observations for individual
sources were reported by \citeauthor{fiedleretal87},
\citet{dennisonetal87}, and \citet{fdjws94}.  

This much larger sample allows us to revisit these issues with a long,
densely-sampled monitoring program.  In \S\ref{sec:observe} we
summarize the observations, in \S\ref{sec:analysis} we combine the new
observations reported here with those from 1979--1988 and present
light curves of the sources and variability analyses employing model
light curve functions, structure functions, and wavelet transforms,
and in \S\ref{sec:conclude} we present our conclusions and
instructions for obtaining these data.

\section{Observations}\label{sec:observe}

The Green Bank Interferometer (GBI) was a facility of the National
Science Foundation and was operated by the National Radio Astronomy
Observatory under contract with the Naval Research Laboratory (NRL)
and the US Naval Observatory (USNO) during the interval 1978~October
to~1996 April~1.  During the interval 1978--1987, light curves of
radio-loud quasars and BL~Lacs were obtained during the Earth Rotation
Parameters Program operation of the instrument
(\citeauthor{fiedleretal87}; \citealt{wfjsfjmm91}).  During
1988--1994.4 the emphasis of the monitoring program was changed to
detect ESEs (\citealt{fdjh87}), and a program of monitoring compact
extragalactic flat-spectrum radio sources for ESEs was instituted.
After~1994 the instrument concentrated on Galactic X-ray binaries and
extragalactic sources of high-energy radiation.  The GBI itself is
described by \citet{hmcw69} and \citet{c73}.  The observing and
calibration procedures relevant to the NRL and USNO observing programs
have been described by \citeauthor{fiedleretal87} and
\citet{wfjsfjmm91}.  Here we shall summarize the details only briefly
and call attention to those procedures which have changed since the
report of \citet{wfjsfjmm91}.

Observations were made on a 2.4~km baseline.  Dual circular
polarization was recorded over a 35~MHz bandwidth at two frequencies
in the S- and X frequency bands.  Until 1989~August (1989.7), the
frequencies were 2.7~GHz (S-band) and~8.1~GHz (X-band); in
1989~September cryogenic receivers were installed, and the frequencies
changed to~2.25~GHz (S-band) and~8.3~GHz (X-band).

Most sources were observed every other day with scans of 10--15~min.\
integration.  Prior to the receiver switch, the observation
frequencies were switched between the two frequency bands every 30~s
within a scan, and the reported flux densities are daily averages of
multiple scans in left-circular polarization only
\citeauthor{fiedleretal87}.  The cryogenic receivers both allow the
two frequencies to be obtained simultaneously and are more sensitive.
After their installation, the observing procedure was changed to
acquire typically only a single scan on a source per day.  Also, the
flux densities reported here for the interval 1988--1994 are the
average of left- and right-circular polarizations.

The calibration was ultimately referred to \objectname[]{3C286}
(\citealt{bgp-tw77}), for which flux densities of 11.85~Jy at~2.25~GHz
(10.53~Jy at~2.7~GHz) and 5.27~Jy at~8.3~GHz (5.34~Jy at~8.1~GHz) were
assumed.  Calibration of the program sources was based on
\objectname[]{0237$-$233} (\objectname[Ohio]{OD$-$263}),
\objectname[]{1245$-$197} (\objectname[Ohio]{ON$-$176.2}),
\objectname[]{1328$+$307} (\objectname[3C]{3C286}), and
\objectname[]{1328$+$254} (\objectname[3C]{3C287}) and was weighted by
the difference in time between the observation of the calibration
source and the program source.  Editing was performed with a running
boxcar mean, and data deviating by more than 10$\sigma$ at X band
or~15$\sigma$ at S band were eliminated.

The uncertainties of the GBI's flux densities measurements are both
flux-density and time dependent.  (The latter occurring at least in part because of the
receiver change in 1989~August.)  \citeauthor{fiedleretal87} presented the
following expressions:
\begin{eqnarray}
\sigma_S^2 & = & (0.037\,\mathrm{Jy})^2 + (0.014 F_S)^2 \nonumber \\
\sigma_X^2 & = & (0.057\,\mathrm{Jy})^2 + (0.049 F_X)^2,
\label{eqn:rmsI}
\end{eqnarray}
with $F_{S,X}$ the S- and X-band flux densities in Jy.  These
expressions are appropriate prior to the installation of cryogenic
receivers in 1989~August.  After the installation of cryogenic
receivers, the appropriate expressions are
\begin{eqnarray}
\sigma_S^2 & = & (0.0037\,\mathrm{Jy})^2 + (0.015 F_S)^2 \nonumber \\
\sigma_X^2 & = & (0.0057\,\mathrm{Jy})^2 + (0.05 F_X)^2.
\label{eqn:rms}
\end{eqnarray}
Because these uncertainties are determined from individual scans, they
also represent a measure of the GBI's short-term instrumental
stability.

Systematic errors are introduced by atmospheric and hardware effects.
By comparing the measured flux densities at various elevation angles
with that expected from an average gain curve, we estimate that these
systematic errors may approach 10\% at S-band and 20\% at X-band
(occasionally higher for extreme local hour angles).  There is also a
systematic ripple, with approximately 5\% peak-to-peak variations,
introduced by the calibration observations of \objectname[3C]{3C286}
(see Figure~\ref{fig:ltcurv} below).  The origin of this ripple is
unknown (\citeauthor{fiedleretal87}), but its 1-yr time scale suggests
a seasonally-induced instrumental effect.

\section{Light Curves and Analyses (Figure~\ref{fig:ltcurv})}\label{sec:analysis}

\begin{figure}[tbh]
 \begin{center}
 \mbox{\psfig{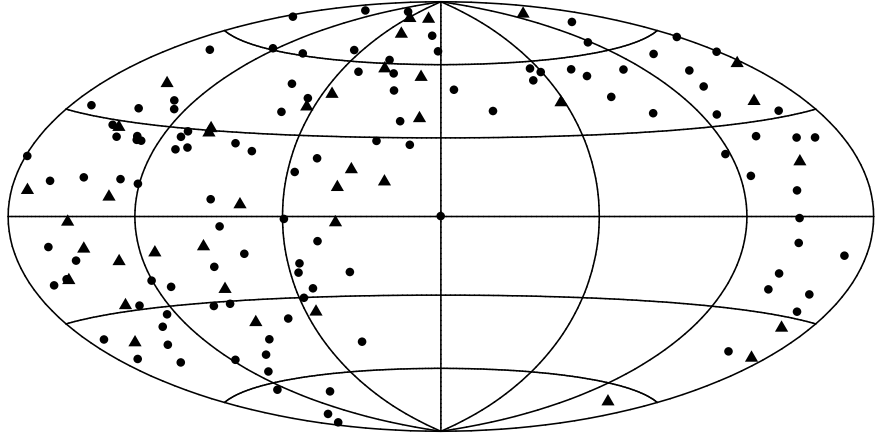}}
 \end{center}
\vspace{-0.5cm}
\caption[]{The distribution on the sky of sources
monitored.  Sources observed during the interval 1988--1996 are
denoted by triangles; sources whose observations commenced prior
to~1988 are denoted by circles.  Sources are plotted using their
Galactic coordinates, and the Galactic center is at the center of the
plot with Galactic longitude increasing to the left.}
\label{fig:sky}
\end{figure}

The list of sources monitored is presented in
Table~\ref{tab:gbisources}, and Figure~\ref{fig:sky} shows their
distribution on the sky in Galactic coordinates.  After the expansion
of the monitoring program for finding ESEs, sources in or near the
Galactic plane were not excluded explicitly.  A selection effect
against low-latitude sources may remain based on the catalogs used
originally to assemble the source list.

\begin{table}[tbh]
\caption{\emph{Available at
\protect\url{http://ese.nrl.navy.mil/GBI/GBI.html}}\label{tab:gbisources}}
\end{table}

Most of the columns in Table~\ref{tab:gbisources} are
self-explanatory.  Columns~(5) and~(6) are the mean flux density over
the interval 1989.7--1994.  We restrict our attention to this interval
because it is the one for which the largest number of sources were
present in the monitoring program and to avoid any systematic effects
from the change of receivers during 1989.  Column~(7) is the structure
index, which is a measure of how compact the source is on
milliarcsecond scales and is defined by \citet{c90}.  A source with a
structure index of~1 is relatively compact, having 90\% or more of its
flux density in a single component, while a structure index of~4
indicates a relatively extended source, typically having two or more
components of approximately equal flux density.  Structure indices are
taken from \citet{fc96} and \citet{fc00}, who determined the structure
indices from the contribution of the sources' structures to
interferometric group delays in VLBI observations.

A comment regarding both the structure index and the visible magnitude
is warranted.  It is well known that sources may show structural
changes on milliarcsecond scales on time scales of decades.  The
structure indices are computed from visibility data obtained after the
conclusion of the GBI monitoring program.  It is plausible that
observations obtained contemporaneously with the GBI program would have
produced a different structure index.  Similarly, many of these sources
are known to be variable at visible wavelengths.  The magnitudes cited
here are the result of a wide variety of observing programs and have
been obtained over more than a decade.  Both the structure index and
$m_V$ should therefore be taken as indicators of a source's structure
and brightness, but future observations could certainly produce
different values of either quantity.

Figure~\ref{fig:ltcurv} presents the light curves of the 149~sources
observed.  While the durations of the light curves are typically
1988--1994, observations of some sources were conducted on a much
longer or much shorter duration.  Of particular note are the 40
sources for which observations began prior to~1988, typically
between~1979 and~1983.  Discontinuities in flux levels at~1989.7
result from the change of observing frequency (\S\ref{sec:observe}).

For various reasons, some sources observed previously
(\citeauthor{fiedleretal87}; \citealt{wfjsfjmm91}) were removed from the
observing schedule.  These sources were \objectname[]{0215$+$015},
\objectname[]{0814$+$425} (\objectname[Ohio]{OJ~425}), \objectname[]{0823$+$493},
\objectname[]{1226$+$023} (\objectname[3C]{3C~273}),
\objectname[]{1345$+$125} (\objectname[4C]{4C~12.50}), and 
\objectname[]{1748$-$253}.  The light curves for 
three sources observed during this interval are not included in this
compilation because they have been published and comprehensive
analyses, including comparison with X-ray observations, have been
performed in previous work.  These sources are
\objectname[]{0236$+$610} (\objectname[]{LSI~$+$61303},
\citealt{rfwtg97}), \objectname[]{GRS~1915$+$105} \citep{fwthzpg96}, and
\objectname[]{2030$+$407} (\objectname[]{Cyg~X-3},
\citealt{wfjg94,wgjffs95,wfpfg96,mccolloughetal99}).

\subsection{Root-Mean-Square Variation}\label{sec:rms}

As our first measure of the variability of these sources (\citeauthor{fiedleretal87}), we use the
rms flux density variation about the mean, defined in the standard
manner as
\begin{equation}
s_F^2 = \frac{1}{N-1}\sum_i [F(t_i) - m_F]^2
\label{eqn:fluxrms}
\end{equation}
for an $N$-point, discretely-sampled flux density time series~$F(t_i)$
having a mean~$m_F$.  The rms flux density incorporates variations on
all time scales, in particular, it incorporates the instrumental
variations described by equation~(\ref{eqn:rms}) in addition to any
intrinsic variations.  In order to avoid a bias from the receiver
change, rms flux densities (and the means used to calculate them) are
determined utilizing only the data from the interval 1989.7--1994,
i.e., after the change in receivers and frequencies.

In order to account for the instrumental contribution, 
we model the rms flux density as being due to at least three terms
\begin{equation}
s_F^2 = s_{\mathrm{GBI,s}}^2 + s_{\mathrm{GBI,l}}^2 + s_i^2.
\label{eqn:rmsmodel}
\end{equation}
Here $s_{\mathrm{GBI,s}}$ is the short-term instrumental variability
(equation~\ref{eqn:rms}), $s_{\mathrm{GBI,l}}$ accounts for any
long-term gain variations or other long-term instrumental effects, and 
$s_i$ is the intrinsic variability.  In what follows we shall analyze
$s_F^\prime \equiv \sqrt{s_F^2 - s_{\mathrm{GBI,s}}^2}$.  Though there 
does appear to be a long-term instrumental or systematic variation
(namely the annual ripple),
this variation does not appear to be flux-density dependent and its
magnitude is generally smaller than those of the two terms on the
right-hand side of equation~(\ref{eqn:rmsmodel}).

\begin{figure}[tbh]
 \begin{center}
 \mbox{\psfig{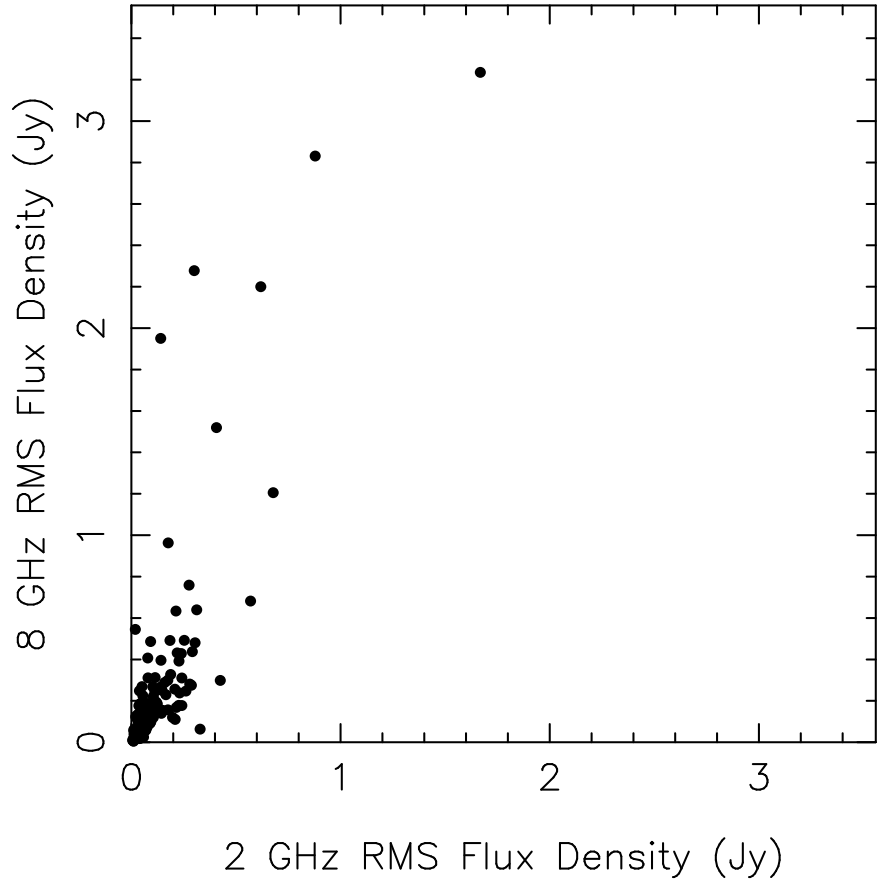}}
 \end{center}
\vspace{-0.5cm}
\caption[]{The 8.3~GHz rms flux densities plotted against the
2.25~GHz rms flux densities.}
\label{fig:rvr}
\end{figure}

Figure~\ref{fig:rvr} shows the rms flux densities from the two
frequency bands plotted against each other.
\citeauthor{fiedleretal87} found a high degree of correlation, $r =
0.97$, between the rms flux densities at the two frequency bands.
They interpreted this correlation as an indication of a common flaring
or variability mechanism in their 36-source sample.  Our larger sample
displays a similarly high degree of correlation, $r = 0.82$.  While
our correlation coefficient is numerically lower than that found by
\citeauthor{fiedleretal87}, our larger sample of sources means that
the correlation is of higher significance.

The slope of the best-fit line for our sample is 2.14.  The
variability spectral index ($s_F \propto \nu^\zeta$) is $\zeta = 0.58$
corresponding to an an electron energy density power-law index of~2.5
for the adiabatic stage of a radio flare in the \cite{mg85} shock
model.  Analyzing a subset of these sources,
\citeauthor{fiedleretal87} found $\zeta = 0.58$ and an electron
density power-law index of~2.4.

\begin{figure}[tbh]
 \begin{center}
 \mbox{\psfig{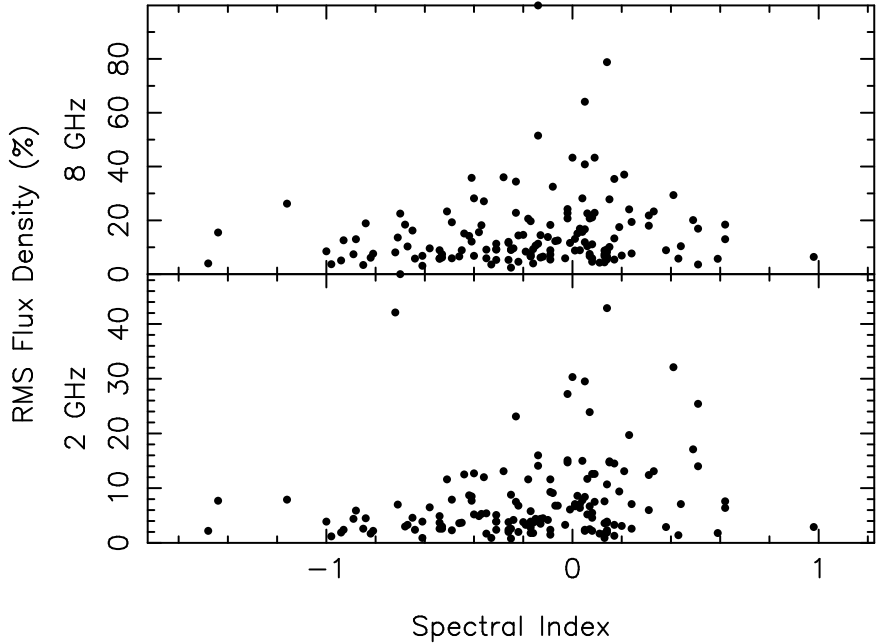}}
 \end{center}
\vspace{-0.5cm}
\caption[]{RMS flux densities as a function of the median spectral
index.
\textit{Top}:~8.3~GHz; 
\textit{Bottom}:~2.25~GHz.}
\label{fig:rvsi}
\end{figure}

\citeauthor{fiedleretal87} also found that flat-spectrum objects tend
to be more variable than steep-spectrum objects.
Figure~\ref{fig:rvsi} illustrates that this trend continues in our
larger sample, though there are more outliers.  However, there appears
to be no other common property among the steep-spectrum, highly
variable sources.  For instance, the three most variable sources in
Figure~\ref{fig:rvsi} with spectral indices $\alpha < -0.5$ are
\objectname[]{1308$+$326} ($\alpha = -0.6$, $s = 49$\%), a relatively
compact quasar; \objectname[]{1845$+$797} ($\alpha = -0.53$, $s =
42$\%), an extended Seyfert~1 galaxy; and \objectname[]{1909$+$048}
($\alpha = -0.88$, $s = 29$\%), a Galactic X-ray binary.

\begin{figure}[tbh]
 \begin{center}
 \mbox{\psfig{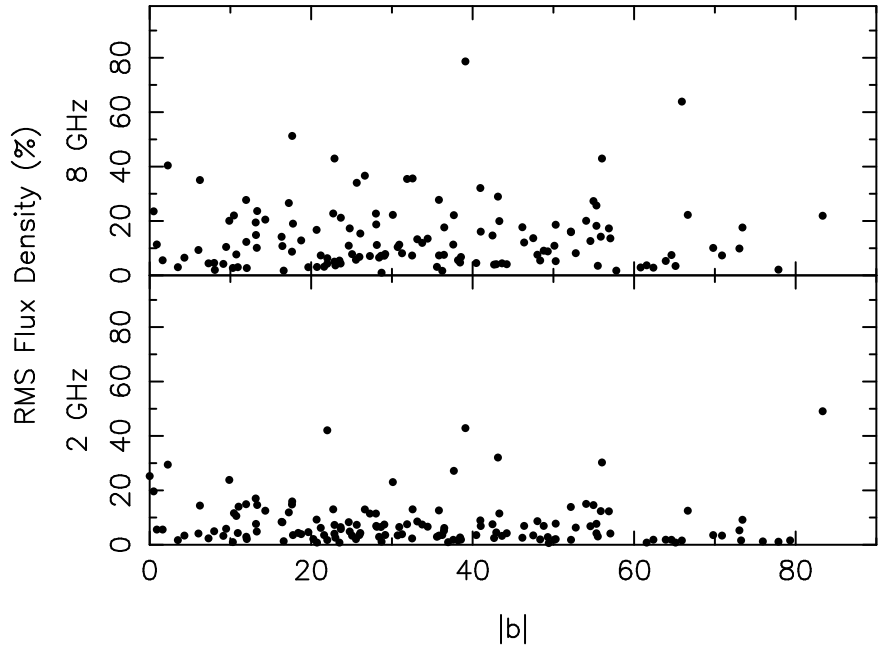}}
 \end{center}
\vspace*{-0.5cm}
\caption[]{RMS flux densities as a function of (absolute) Galactic latitude.
\textit{Top}:~8.3~GHz;
\textit{Bottom}:~2.25~GHz.}
\label{fig:rvb}
\end{figure}

Because flat-spectrum objects are generally more compact than
steep-spectrum objects \citep{ko88}, an alternate explanation of, or
at least additional contribution to, their variability might be
interstellar scintillation.  As we discuss below (\S\ref{sec:sf}), at
least a fraction of these sources' variability does arise from
interstellar scintillation.  Interstellar scintillation will be shown
to be important on time scales of order 10~days.  However, the rms
flux density measures variability on all time scales and tends to be
dominated by longer time scale variations, so scintillation is largely
unimportant.  Figure~\ref{fig:rvb} shows that there is no correlation
between a sources's Galactic latitude and its rms flux density; the
correlation coefficient between rms flux density and Galactic latitude
is similar for the two frequencies, $r \simeq -0.05$, and not
significant.  In any simple model for the contribution of interstellar
scintillation to source variability, the contribution should increase
for sources closer to the Galactic plane.  We regard
Figure~\ref{fig:rvb} as an indication that scintillation is not a
dominant contribution to the rms variability.

We regard Figures~\ref{fig:rvr} and~\ref{fig:rvsi} as evidence of a
common flaring or variability mechanism in these sources.  First, the
8.3~GHz variability is larger in magnitude than the 2.25~GHz variability.
This is consistent with various models in which outbursts or flares
are reduced in magnitude (and delayed in time) at low frequencies
relative to high frequencies.  Second, flat-spectrum sources tend to
be more variable than steep-spectrum sources.

While the results of this variability analysis are consistent with the
scenario that all sources have a common emission mechanism, we caution
that there could be a host of unmodeled systematic effects that could
mask true variability or conspire to produce apparent variability.
(We regard this as a general difficulty with these kind of analyses,
not just with the data from this instrument.)  Among the possible
systematic effects are elevation and hour angle effects and the
sidelobe pattern on the sky.  While we see no indications of such
effects in these data, others who intend to make use of these data
should bear such possibilities in mind.

\subsection{Light Curve Classification}\label{sec:arima}

\citeauthor{fiedleretal87} conducted an autocorrelation analysis of
the light curves presented therein.  From the autocorrelation
functions (ACFs) they extracted decorrelation time scales.  They found
strong correlations between the decorrelation time scales at the two
frequencies and with the spectral indices of the sources.  This ACF
analysis in \citeauthor{fiedleretal87} contributed, in part, to their
conclusion that the sources likely had a common emission mechanism.

A key assumption in calculating an ACF is that the underlying time
series is in some sense stationary.  In extending
\citeauthor{fiedleretal87}'s analysis to our larger sample, we have
concluded that the assumption of stationarity is not warranted for
these light curves and that an ACF analysis can be misleading.  One
indication that these time series are, in general, not stationary can
be obtained from Figure~\ref{fig:ltcurv}.  Some of the sources exhibit
trends of increasing or decreasing flux density, indicating that that
the mean flux density may not be a well-defined quantity and that the
time series is not stationary, at least over the length of the time
series.  We also found that by choosing different sub-intervals of a
light curve---for those sources common to this work and
\citeauthor{fiedleretal87}---the qualitative shape and the inferred
decorrelation time scale of the ACF could change dramatically, another
symptom of non-stationarity.

In their analysis \cite{haa91} employed the first-order structure
function to determine the time scale and behavior of their sample of
sources, with the time behavior of the sources quantified by means of
the logarithmic structure function slope.  We defer use of the
structure function to \S\ref{sec:sf} and employ an alternate means for
describing our time series, for two reasons.  First, \cite{haa91} find
a broad distribution of time scales in their sample.  Fully 30\% of
their sample has a time scale comparable to or longer than the 6-yr
duration for which most of this sample has been monitored.  A similar
structure function analysis of this monitoring program would
presumably yield incomplete or biased estimates of the structure
function and characteristic time scales.  Second, while a relatively simple magnetohydrodynamic
model describes the time series of the total and polarized intensity
of \objectname[]{BL~Lac}, \objectname[]{3C~279}, and
\objectname[]{OT~081} \citep{haa89a,haa89b,haa91}, there are important
differences between the model parameters for the three sources.
At least in the context of this class of models, a time series
analysis might identify those sources having extreme model parameters
or those sources for which the model was not applicable.

As an alternate means of describing these time series, we chose to
employ autoregressive, integrated, moving average (ARIMA) models.
ARIMA models have found a variety of uses in both describing and
forecasting time series for industrial and economic applications
\citep{bjrj94} and limited use in astronomy
\citep{s81,o87,s90,kl93,pkws98}.  ARIMA models also have the desirable
property that they are not limited to modeling stationary processes.

Our objective is two-fold.  First, we assess the extent to which ARIMA 
models can describe the light curves and to what extent our sample
suggests that a common emission process exists for these sources.
Second, we wish to identify if there are any sources for which the
existing emission models might not be applicable or for which the
model parameters might be required to be extreme.

Because of the limited use of ARIMA models in astronomy and for
notational purposes, we review the basic ARIMA model.  In a $p$-order
AR process the flux value at the current time interval~$F_i$ depends
upon its values at the $p$ previous time increments and a white noise
process~$f_i$:
\begin{equation}
F_i = f_i + \phi_1F_{i-1} + \phi_2F_{i-2} + \ldots + \phi_pF_{i-p},
\label{eqn:ar}
\end{equation}
where the $\phi_j$ are coefficients.  In a $q$-order MA process the
flux value at the current time depends upon the values of a white
noise process at the current time and $q$ previous time increments
\begin{equation}
F_i = f_i - \theta_1f_{i-1} - \theta_2f_{i-2} - \ldots - \theta_qf_{i-q},
\label{eqn:ma}
\end{equation}
where $\theta_j$ are coefficients.  A $d$-order integrated process is
one in which the value at the current and $d$ previous time increments
is driven by a white noise process
\begin{equation}
F_i + (-1)^1{d\choose1}F_{i-1} + (-1)^2{d\choose2}F_{i-2} + \ldots 
 + (-1)^dF_{i-d} = f_i.
\label{eqn:diff}
\end{equation}
Here ${d\choose k}$ is the binomial coefficient.  A $(p, d, q)$ ARIMA
process combines these individual mechanisms.  Relevant to the shocked
jet flow model \citep{mg85,haa89a} is the white noise process~$f_i$ in
equations~(\ref{eqn:ar}), (\ref{eqn:ma}), and~(\ref{eqn:diff}).  This
white noise process can be considered to be a series of shocks driving
the system.  An ARIMA process allows the flux density at the current
time to depend upon its value and the value of the white noise process
at previous times, as might be expected if the emission from these
sources arises from multiple, evolving shocks.

We followed analysis procedures described by \cite{bjrj94}: The
various processes can be identified by comparing the behavior of their
ACFs and \emph{partial autocorrelation functions}.  The partial
autocorrelation function measures the correlation between~$F_i$
and~$F_{i-k}$ taking into account the intermediate values~$F_{i-1}$,
$F_{i-2}$, \ldots, $F_{i-k+1}$.  Key to our classification of these
light curves are the following properties of the ACF and the partial
autocorrelation function: For an AR($p$) process the ACF cuts off
after the first $p$ lags while its partial autocorrelation function
tails off exponentially after $p$ lags; conversely, for an MA($q$)
process the ACF tails off exponentially after $q$ lags while its
partial autocorrelation function cuts off after the first $q$ lags.  A
$d$-order integrated process can be identified by taking $d$
differences of the time series (equation~\ref{eqn:diff}) before
forming the ACFs and partial autocorrelation functions.  In evaluating
whether a partial autocorrelation function or ACF has cut off, we took
the fractional uncertainty in estimating both the ACF and partial
autocorrelation function to be of order $N^{-1/2}$ for an N-point
light curve.  In order to minimize any potential effects arising from
the receiver switch in~1989, we restricted the analysis to the
interval 1989.7--1994.

Earlier we indicated that we found the ACF alone to be an unreliable
means of characterizing these light curves.  The use of the ACF here
does not contradict those earlier statements.  First, the difficulty
cited with the ACF was due to the non-stationary nature of the light
curves.  The differencing (equation~\ref{eqn:diff}) is designed
precisely to reduce the differenced light curve to a stationary, or
nearly stationary, time series.  Second, the ACF is used here in
conjunction with the partial autocorrelation function which is
calculated independently and by a different means than the
\hbox{ACF}.  Thus, even if the ACF is affected by the non-stationarity 
of the light curve, the partial autocorrelation function will not be.

We have focussed on ARIMA processes with orders $p$, $d$, $q \le 2$.
We have imposed this limit for two reasons.  First, low-order ARIMA
models have had considerable success in describing many other kinds of
time series \citep{bjrj94}.  Second, our focus is on establishing
whether a common emission process(es) can describe the light curves.
Higher-order ARIMA models may also produce acceptable descriptions of
the light curves, though with increasing ARIMA order the reality of
the model must be increasingly questioned.  Furthermore, if a source
cannot be described by any of these low-order ARIMA models that may be 
an important clue regarding the applicability of the shocked jet model 
to that source.

Our description of these light curves in terms of ARIMA models is
necessarily not unique.  First, there is the restriction on the order
of the ARIMA models considered, though as noted above high-order ARIMA
models may not be realistic.  Second, even within our restricted set
of models, the solutions may not be unique.  For instance, as
\cite{bjrj94} discuss, the boundary between a stationary and
non-stationary process is not distinct.  A stationary time series may
appear non-stationary if the duration over which it is sampled is too
short (and conversely for a non-stationary time series appearing
stationary).  Finally, there may be other models to describe these
light curves besides the class of ARIMA models.  Our objective,
though, is not to provide a unique description of each light curve,
but to assess the extent to which the different sources may have
similar emission mechanisms.  For this purpose, the description in
terms of ARIMA models is sufficient.

A key assumption in the description (and analysis methods) of ARIMA
models is that the time series is sampled uniformly.  We do not have
uniformly-sampled time series, though they are often nearly uniformly
sampled.  We have taken two steps to transform these light curves into
uniformly sampled time series.  First, we have gridded the existing
data onto a uniform grid.  As the typical spacing between samples is
within a few percent of 2~days after~1988, this step should produce
little effect on our model identification.  Second, we have linearly
interpolated between missing data.  We have verified in the following
manner that this step does not impact adversely our model
identification.  We took uniformly-sampled time series (series~B
and~C) from the examples provided by \cite{bjrj94}.  We then produced
non-uniformly-sampled time series by using the sampling from one of
the sources observed, i.e., we introduced gaps whose location and
duration were taken from one of the observed light curves.  Our model
identification procedure nonetheless identified the same models as
\cite{bjrj94} cite for the uniformly-sampled time series.

Tables~\ref{tab:arima2} and~\ref{tab:arima8} show the model(s)
identified for each source at the two frequencies.
Table~\ref{tab:noarima} lists those sources for which no model was
identified.

\begin{deluxetable}{lcccccccccccc}
\tabletypesize{\scriptsize}
\tablewidth{0pt}
\tablecaption{ARIMA Model Identifications at~2.25~GHz\label{tab:arima2}}
\tablehead{\colhead{Source}&
	\colhead{(1 0 0)}&
	\colhead{(1 0 1)}&
	\colhead{(2 0 0)}&
	\colhead{(2 0 2)}&
	\colhead{(0 1 1)}&
	\colhead{(1 1 0)}&
	\colhead{(1 1 1)}&
	\colhead{(2 1 0)}&
	\colhead{(2 1 2)}&
	\colhead{(0 2 1)}&
	\colhead{(1 2 1)}&
	\colhead{(2 2 2)}}
\startdata
0003$+$380 &   &   &   & X &   &   &   &   &   & X &   &  \\ 
0003$-$066 &   &   &   &   & X &   & X &   &   & X &   & X\\ 
0016$+$731 &   &   &   &   & X &   &   & X &   & X &   &  \\ 
0019$+$058 &   &   &   &   & X &   & X &   &   & X &   & X\\ 
0035$+$121 &   &   &   &   &   &   &   &   &   & X &   & X\\ 
0035$+$413 &   &   &   &   &   &   &   & X &   & X &   & X\\ 
0055$+$300 &   &   &   &   & X &   & X &   &   &   &   & X\\ 
0056$-$001 &   &   &   &   & X &   & X &   &   &   &   &  \\ 
							       
\\							       
							       
0113$-$118 &   &   &   &   &   &   & X &   &   &   &   &  \\ 
0123$+$257 & X & X &   &   & X &   & X &   &   & X & X &  \\ 
0130$-$171 & X & X &   &   & X & X & X &   &   & X & X &  \\ 
0133$+$476 &   &   &   &   & X & X & X &   &   & X &   & X\\ 
0134$+$329 &   &   &   &   & X &   &   &   & X &   &   &  \\ 
0147$+$187 &   &   &   &   & X &   & X &   &   &   &   &  \\ 
							       
\\							       
							       
0201$+$113 & X &   &   &   &   &   &   & X &   & X & X &  \\ 
0202$+$319 &   &   &   &   &   &   &   &   &   & X &   & X\\ 
0212$+$735 &   &   &   &   &   &   & X &   &   &   &   &  \\ 
0224$+$671 &   &   &   &   & X &   &   & X & X & X &   & X\\ 
0235$+$164 &   &   &   &   & X &   &   & X & X & X &   & X\\ 
0237$-$233 &   &   &   &   & X &   &   &   &   &   &   & X\\ 
0256$+$075 &   &   &   &   & X &   &   &   &   & X &   &  \\ 
							       
\\							       
							       
0300$+$470 &   &   &   &   & X & X &   &   & X &   &   &  \\ 
0316$+$413 &   &   &   &   & X &   &   &   & X &   &   & X\\ 
0319$+$121 &   &   &   &   & X &   & X &   &   &   &   &  \\ 
0333$+$321 &   &   &   &   & X &   &   &   &   & X &   &  \\ 
0336$-$019 &   &   &   &   &   &   & X &   &   & X &   &  \\ 
0337$+$319 &   &   &   &   & X &   &   &   & X &   &   & X\\ 
0355$+$508 &   &   &   &   & X &   &   &   & X & X &   & X\\ 
							       
\\							       
							       
0400$+$258 &   &   &   &   &   &   & X &   &   &   &   &  \\ 
0403$-$132 &   &   &   &   &   &   & X &   &   & X &   & X\\ 
0415$+$379 &   &   &   &   & X &   &   & X &   & X &   &  \\ 
0420$-$014 &   &   &   &   &   &   & X &   &   & X &   & X\\ 
0440$-$003 &   &   &   &   & X &   &   &   &   &   &   &  \\ 
0444$+$634 &   &   &   &   & X & X &   &   & X & X &   & X\\ 
0454$+$844 &   &   &   &   &   &   & X &   &   & X &   & X\\ 
							       
\\							       
							       
0500$+$019 &   &   &   &   & X &   & X &   &   & X &   & X\\ 
0528$+$134 &   &   &   &   &   & X & X &   &   &   &   & X\\ 
0532$+$826 &   &   &   &   &   &   & X &   &   & X & X &  \\ 
0537$-$158 &   &   &   &   & X &   &   & X &   & X &   & X\\ 
0538$+$498 &   &   &   &   &   &   & X &   &   & X &   & X\\ 
0552$+$398 &   &   &   &   &   &   &   &   &   &   &   & X\\ 
0555$-$132 &   &   &   &   & X &   &   &   &   & X &   & X\\ 
							       
\\							       
							       
0615$+$820 &   &   &   &   &   &   &   &   & X &   &   &  \\ 
0624$-$058 &   &   &   &   & X &   &   &   & X &   &   & X\\ 
0633$+$734 &   &   & X &   &   &   &   &   &   &   &   &  \\ 
0650$+$371 &   &   &   &   &   &   & X &   &   & X &   & X\\ 
0653$-$033 &   &   &   &   & X & X & X &   &   & X &   & X\\ 
							       
\\							       
							       
0716$+$714 &   &   &   &   &   &   &   &   & X & X & X &  \\ 
0723$+$679 &   & X &   &   & X &   &   & X & X & X &   &  \\ 
0723$-$008 &   &   &   &   & X &   &   &   &   & X &   & X\\ 
0742$+$103 &   &   & X &   &   &   &   &   &   &   &   &  \\ 
0743$+$259 &   &   &   &   & X &   & X &   &   & X &   & X\\ 
0759$+$183 &   & X & X &   &   &   &   &   &   & X & X &  \\ 
							       
\\							  
							       
0804$+$499 &   &   &   &   &   &   &   &   & X & X &   & X\\ 
0818$-$128 &   &   &   &   &   &   & X &   &   & X &   & X\\ 
0827$+$243 &   &   &   &   &   &   &   &   &   & X &   & X\\ 
0837$+$035 &   & X &   &   & X &   &   &   &   & X &   &  \\ 
0851$+$202 &   &   &   &   &   &   & X &   &   & X &   &  \\ 
0859$-$140 &   &   &   &   & X &   &   &   &   & X &   & X\\ 
							       
\\							       
							       
0922$+$005 &   &   &   &   & X &   & X & X &   & X &   & X\\ 
0923$+$392 &   &   &   &   & X &   & X &   &   &   &   &  \\ 
0938$+$119 &   &   &   &   &   &   &   &   &   & X &   &  \\ 
0945$+$408 &   &   &   &   & X &   &   &   &   &   &   &  \\ 
0952$+$179 & X &   &   & X &   &   & X &   &   & X &   &  \\ 
							       
\\							       
							       
1020$+$400 &   &   &   &   & X &   &   &   & X & X &   &  \\ 
1022$+$194 &   &   &   &   & X &   & X &   &   & X &   & X\\ 
1036$-$154 &   &   &   &   & X &   & X &   &   & X &   & X\\ 
1038$+$528 &   &   &   &   &   &   & X &   &   & X & X &  \\ 
1055$+$018 &   &   &   &   &   &   &   &   &   & X &   &  \\ 
							       
\\							       
							       
1100$+$772 &   & X &   &   & X &   &   &   &   &   & X &  \\ 
1116$+$128 &   &   &   &   &   &   &   &   &   & X & X &  \\ 
1123$+$264 &   &   &   &   & X &   &   &   & X &   &   & X\\ 
1127$-$145 &   &   &   &   &   &   &   &   &   & X &   & X\\ 
1128$+$385 &   &   &   & X &   &   &   &   & X & X &   &  \\ 
1145$-$071 &   &   &   &   & X &   & X &   &   & X &   &  \\ 
1150$+$812 &   &   &   &   &   &   &   &   &   & X &   & X\\ 
1155$+$251 &   &   &   &   &   &   & X &   &   & X &   & X\\ 
							       
\\							       
							       
1200$-$051 &   &   &   &   &   &   & X &   &   & X &   & X\\ 
1243$-$072 & X &   &   & X &   &   &   &   &   & X &   &  \\ 
1245$-$197 &   &   & X & X & X &   &   &   & X &   &   & X\\ 
1250$+$568 &   & X &   &   & X &   &   &   &   & X &   & X\\ 
1253$-$055 &   &   &   & X &   &   &   &   &   &   &   &  \\ 
							       
\\							       
							       
1302$-$102 &   &   &   &   & X &   &   &   &   &   &   &  \\ 
1308$+$326 &   &   &   &   &   &   & X &   &   & X &   & X\\ 
1328$+$307 &   &   &   &   & X &   &   &   & X &   &   & X\\ 
1354$+$195 &   &   &   &   &   &   &   &   &   & X &   &  \\ 
							       
\\							       
							       
1404$+$286 &   &   &   & X &   &   &   &   &   & X &   & X\\ 
1409$+$524 &   &   &   &   & X &   &   &   & X &   &   &  \\ 
1413$+$135 &   &   &   &   &   &   &   &   & X & X & X &  \\ 
1430$-$155 & X &   &   &   & X &   & X & X &   & X &   & X\\ 
1449$-$012 &   &   &   &   &   &   &   &   &   & X &   & X\\ 
							       
\\							       
							       
1502$+$106 &   &   &   &   & X &   &   &   &   & X &   & X\\ 
1511$+$238 &   &   &   &   & X &   &   &   &   & X &   & X\\ 
1514$+$197 &   &   &   &   &   & X &   &   &   &   &   &  \\ 
1525$+$314 &   &   &   &   & X & X & X &   &   & X &   & X\\ 
1538$+$149 &   &   &   &   & X & X & X &   &   & X &   & X\\ 
1555$+$001 &   &   &   &   & X &   & X & X &   &   &   &  \\ 
							       
\\							       
							       
1614$+$051 &   &   &   &   &   &   &   & X &   &   &   &  \\ 
1624$+$416 &   &   &   &   & X &   &   &   & X &   &   & X\\ 
1635$-$035 &   &   &   &   & X & X & X &   &   & X &   & X\\ 
1641$+$399 &   &   &   &   &   &   & X &   &   & X &   &  \\ 
1655$+$077 &   &   &   &   & X &   &   &   & X & X &   & X\\ 
1656$+$477 &   & X & X &   & X &   & X & X &   &   &   & X\\ 
							       
\\							       
							       
1741$-$038 &   &   &   &   & X & X &   &   & X & X &   &  \\ 
1749$+$096 &   &   &   &   &   & X &   &   &   &   &   &  \\ 
1749$+$701 &   &   &   &   & X &   & X &   &   & X &   &  \\ 
1756$+$237 &   &   & X &   & X &   &   &   & X &   &   & X\\ 
							       
\\							       
							       
1803$+$784 &   &   &   &   & X &   &   &   &   & X &   &  \\ 
1807$+$698 &   &   &   &   &   &   &   &   & X &   &   &  \\ 
1821$+$107 &   &   &   &   & X &   &   &   & X & X &   & X\\ 
1823$+$568 & X &   &   & X &   &   & X &   &   & X &   & X\\ 
1828$+$487 &   &   &   &   & X &   & X &   &   &   &   & X\\ 
1830$+$285 &   &   &   &   & X & X &   &   & X & X &   &  \\ 
1845$+$797 &   &   &   &   & X &   & X &   &   &   & X &  \\ 
							       
\\							       
							       
1909$+$048 &   &   &   &   &   &   &   & X & X &   &   &  \\ 
1928$+$738 &   &   &   &   & X &   & X &   &   & X &   & X\\ 
1943$+$228 &   &   &   &   & X &   &   &   &   &   &   & X\\ 
1947$+$079 &   &   &   &   &   &   &   &   &   &   &   & X\\ 
							       
\\							       
							       
2005$+$403 &   &   &   &   & X &   &   &   & X & X &   & X\\ 
2008$-$068 & X &   &   & X & X &   & X &   &   & X &   &  \\ 
2032$+$107 &   &   &   &   & X &   & X &   &   & X &   &  \\ 
2037$+$511 &   &   &   &   & X &   &   &   &   &   &   & X\\ 
2047$+$098 &   &   &   &   &   &   &   &   &   &   &   & X\\ 
2059$+$034 &   &   &   &   &   &   & X & X &   & X & X &  \\ 
							       
\\							       
							       
2105$+$420 &   &   &   &   & X &   &   &   &   &   &   &  \\ 
2113$+$293 &   &   &   &   & X & X & X &   &   & X &   & X\\ 
2121$+$053 &   &   &   &   &   & X &   &   &   & X &   & X\\ 
2134$+$004 &   &   &   &   &   &   & X &   &   & X &   & X\\ 
2155$-$152 &   &   &   &   & X & X &   &   & X &   &   & X\\ 
							       
\\							       
							       
2200$+$420 &   &   &   &   &   &   & X &   &   &   &   &  \\ 
2209$+$081 &   &   &   &   & X &   &   &   &   &   &   & X\\ 
2214$+$350 &   &   &   &   &   &   &   &   &   & X &   & X\\ 
2251$+$244 &   &   &   &   & X &   &   &   & X &   &   & X\\ 
							       
\\							       
							       
2319$+$272 &   &   &   &   &   &   & X &   &   &   & X &  \\ 
2344$+$092 &   &   &   &   & X &   &   &   &   &   &   & X\\ 

\enddata
\tablecomments{Only sources for which at least one ARIMA model was
identified are shown, and only ARIMA models that describe at least one 
source are listed.}
\end{deluxetable}

\begin{deluxetable}{lcccccccccccc}
\tabletypesize{\scriptsize}
\tablewidth{0pt}
\tablecaption{ARIMA Model Identifications at~8.3~GHz\label{tab:arima8}}
\tablehead{
	\colhead{Source}&
	\colhead{(1 0 0)}&
	\colhead{(1 0 1)}&
	\colhead{(2 0 0)}&
	\colhead{(2 0 2)}&
	\colhead{(0 1 1)}&
	\colhead{(1 1 0)}&
	\colhead{(1 1 1)}&
	\colhead{(2 1 0)}&
	\colhead{(2 1 2)}&
	\colhead{(0 2 1)}&
	\colhead{(1 2 1)}&
	\colhead{(2 2 2)}}
\startdata
0003$+$380 &   &   &   &   &   &   &   &   & X & X &   & X \\ 
0003$-$066 &   &   &   &   & X &   &   &   & X &   &   &   \\ 
0016$+$731 &   &   &   &   & X &   &   &   &   &   &   &   \\ 
0019$+$058 &   &   &   &   & X &   &   & X & X & X &   & X \\ 
0035$+$121 &   &   &   &   & X &   &   &   & X &   &   &   \\ 
0055$+$300 &   &   &   &   & X &   & X &   &   & X &   & X \\ 
0056$-$001 &   &   &   &   & X &   & X &   &   & X &   & X \\ 
								
\\								
								
0113$-$118 &   &   &   &   &   &   & X &   &   & X &   &   \\ 
0123$+$257 & X & X &   &   & X &   & X &   &   & X &   & X \\ 
0130$-$171 &   &   &   &   & X &   & X & X &   &   &   &   \\ 
0133$+$476 &   &   &   &   & X &   &   &   &   &   &   &   \\ 
0134$+$329 &   & X &   &   & X &   &   &   & X &   &   & X \\ 
0147$+$187 &   &   &   &   & X &   &   &   &   &   &   &   \\ 
								
\\								
								
0201$+$113 &   &   &   &   &   &   &   &   &   & X &   &   \\ 
0202$+$319 &   &   &   &   &   &   &   &   & X & X &   &   \\ 
0212$+$735 &   &   &   &   & X &   &   &   & X &   &   &   \\ 
0224$+$671 &   &   &   &   & X &   &   &   &   & X &   &   \\ 
0235$+$164 &   &   &   &   &   &   &   & X &   & X &   & X \\ 
0237$-$233 &   &   &   &   &   &   &   &   &   & X &   & X \\ 
0256$+$075 &   &   &   &   & X &   & X &   &   &   &   & X \\ 
								
\\								
								
0300$+$470 &   &   &   &   &   &   &   &   &   &   &   & X \\ 
0316$+$413 &   &   &   &   &   &   & X &   &   & X &   & X \\ 
0319$+$121 &   & X &   &   &   &   & X &   &   &   &   &   \\ 
0333$+$321 &   &   &   &   & X &   &   &   &   &   &   & X \\ 
0336$-$019 &   &   &   &   & X &   &   &   & X &   &   &   \\ 
0337$+$319 &   & X &   &   & X &   &   &   &   & X &   & X \\ 
0355$+$508 &   &   &   &   & X &   &   &   &   & X &   & X \\ 
								
\\								
								
0400$+$258 &   &   &   &   &   &   &   &   &   & X &   &   \\ 
0403$-$132 &   &   &   &   &   &   &   &   &   & X &   & X \\ 
0415$+$379 &   &   &   &   & X &   &   &   &   & X &   & X \\ 
0420$-$014 &   &   &   &   &   &   & X &   &   & X &   & X \\ 
0440$-$003 &   &   &   &   &   &   &   &   & X &   &   &   \\ 
0444$+$634 &   &   &   &   & X &   &   &   & X & X &   & X \\ 
0454$+$844 &   &   &   &   & X &   &   &   & X & X &   & X \\ 
								
\\								
								
0500$+$019 &   &   &   &   & X &   &   &   & X & X &   & X \\ 
0532$+$826 &   &   &   &   &   &   &   &   &   & X &   &   \\ 
0537$-$158 &   &   &   &   & X &   &   &   &   & X &   &   \\ 
0538$+$498 &   &   &   & X &   &   &   &   & X & X &   & X \\ 
0552$+$398 &   &   &   &   &   &   &   &   & X & X &   & X \\ 
0555$-$132 &   &   &   &   & X &   &   &   & X &   &   & X \\ 
								
\\								
								
0615$+$820 &   &   &   &   & X &   &   &   &   & X &   &   \\ 
0624$-$058 &   & X &   &   & X &   &   &   & X &   &   & X \\ 
0633$+$734 &   &   &   &   & X &   &   &   &   &   &   &   \\ 
0650$+$371 &   &   &   &   & X &   & X &   &   &   &   & X \\ 
0653$-$033 &   &   &   &   & X &   &   &   & X &   &   & X \\ 
								
\\								
								
0716$+$714 &   &   &   &   & X &   &   &   &   & X &   &   \\ 
0723$+$679 &   &   &   &   & X &   &   &   & X &   &   &   \\ 
0723$-$008 &   &   &   &   & X &   & X &   &   &   &   &   \\ 
0742$+$103 &   & X & X &   & X &   &   &   & X &   &   & X \\ 
0743$+$259 &   &   &   &   & X &   &   &   & X & X &   & X \\ 
0759$+$183 &   &   &   &   & X &   &   &   & X &   &   & X \\ 
								
\\								
								
0804$+$499 &   &   &   &   & X &   &   &   & X & X &   & X \\ 
0818$-$128 &   &   &   &   &   &   &   &   &   &   &   & X \\ 
0827$+$243 &   &   &   &   & X &   &   &   & X &   &   & X \\ 
0836$+$710 &   &   &   &   & X &   &   &   & X &   &   & X \\ 
0837$+$035 &   &   &   &   & X &   &   &   &   & X &   &   \\ 
0851$+$202 &   &   &   &   & X &   &   &   & X & X &   &   \\ 
0859$-$140 &   &   &   &   & X &   &   &   & X &   &   &   \\ 
								
\\								
								
0922$+$005 &   &   &   &   &   &   &   &   & X & X &   & X \\ 
0923$+$392 &   &   &   &   & X &   &   &   & X &   &   &   \\ 
0938$+$119 & X &   &   & X &   &   &   &   &   &   &   &   \\ 
0945$+$408 &   &   &   &   &   &   &   &   & X & X &   & X \\ 
0952$+$179 &   &   &   &   & X &   &   &   & X &   &   &   \\ 
0954$+$658 &   &   &   &   &   &   &   &   &   & X &   &   \\ 
								
\\								
								
1020$+$400 &   &   &   &   & X &   &   &   &   & X &   & X \\ 
1022$+$194 &   &   &   &   & X &   &   &   & X &   &   &   \\ 
1036$-$154 &   &   &   &   & X &   &   &   & X &   &   &   \\ 
1038$+$528 &   &   &   &   & X &   &   &   & X & X &   &   \\ 
1055$+$018 &   &   &   &   & X &   &   &   & X &   &   & X \\ 
								
\\								
								
1100$+$772 &   &   &   &   &   &   &   &   &   & X &   &   \\ 
1116$+$128 &   &   &   &   & X &   &   &   & X &   &   & X \\ 
1123$+$264 &   &   &   &   &   &   &   &   &   & X &   & X \\ 
1127$-$145 &   &   &   &   & X &   &   &   & X &   &   &   \\ 
1128$+$385 &   &   &   &   &   &   &   &   &   & X & X &   \\ 
1145$-$071 &   &   &   &   &   &   & X &   &   & X &   &   \\ 
1150$+$812 &   &   &   &   & X &   &   &   &   & X &   & X \\ 
1155$+$251 &   &   &   &   & X &   & X &   &   & X &   & X \\ 
								
\\								
								
1200$-$051 &   &   &   &   & X &   &   &   & X &   &   &   \\ 
1225$+$368 &   &   &   &   &   &   &   &   &   & X &   &   \\ 
1243$-$072 &   &   &   &   & X &   &   &   &   &   &   & X \\ 
1245$-$197 &   &   &   &   & X &   &   &   & X &   &   &   \\ 
1250$+$568 & X &   &   &   & X &   &   &   &   & X &   & X \\ 
1253$-$055 &   &   &   &   &   &   &   &   &   &   & X &   \\ 
								
\\								
								
1302$-$102 &   &   &   &   & X &   & X &   &   &   &   & X \\ 
1308$+$326 &   &   &   &   &   &   &   &   & X & X &   &   \\ 
1328$+$307 &   &   &   &   &   &   & X &   &   & X &   &   \\ 
1354$+$195 &   &   &   &   & X &   & X &   &   &   &   & X \\ 
								
\\								
								
1404$+$286 &   &   & X &   & X &   & X &   &   &   &   &   \\ 
1409$+$524 &   &   &   &   & X &   &   &   &   &   &   & X \\ 
1413$+$135 &   &   &   &   & X &   &   &   & X &   &   &   \\ 
1430$-$155 &   &   &   &   &   &   &   &   &   & X &   &   \\ 
1438$+$385 &   &   &   &   &   &   & X &   &   & X &   & X \\ 
1449$-$012 &   &   &   &   & X &   &   &   & X &   &   &   \\ 
1455$+$247 & X &   &   & X &   &   & X &   &   & X &   & X \\ 
								
\\								
								
1502$+$106 &   &   &   &   & X &   &   &   &   &   &   & X \\ 
1511$+$238 & X &   &   &   &   &   &   &   &   & X &   & X \\ 
1514$+$197 &   &   &   &   &   &   & X &   &   & X &   &   \\ 
1525$+$314 & X &   &   & X &   &   &   &   &   &   & X &   \\ 
1538$+$149 &   &   &   &   & X &   &   &   & X & X &   & X \\ 
1555$+$001 &   &   &   &   & X &   &   &   &   &   &   &   \\ 
								
\\								
								
1611$+$343 &   &   &   &   &   &   &   &   & X & X &   & X \\ 
1614$+$051 &   &   &   &   & X &   &   &   &   &   &   &   \\ 
1624$+$416 &   &   &   &   &   &   &   &   & X &   &   &   \\ 
1635$-$035 &   &   &   &   & X &   &   &   & X &   &   &   \\ 
1641$+$399 &   &   &   &   &   &   & X &   &   & X &   & X \\ 
1655$+$077 &   &   &   &   &   &   &   &   &   & X &   & X \\ 
1656$+$477 &   &   & X & X &   &   & X &   &   &   &   &   \\ 
								
\\								
								
1741$-$038 &   &   &   &   &   &   & X &   &   &   &   &   \\ 
1749$+$096 &   &   &   &   &   &   & X &   &   &   &   & X \\ 
1749$+$701 &   &   &   &   &   &   &   &   &   & X &   &   \\ 
1756$+$237 &   &   &   &   & X &   &   &   &   &   &   & X \\ 
								
\\								
								
1803$+$784 &   &   &   &   &   &   & X &   &   & X &   &   \\ 
1807$+$698 &   &   &   &   & X &   &   &   &   & X &   & X \\ 
1821$+$107 &   &   &   & X & X &   &   &   & X & X &   & X \\ 
1823$+$568 &   &   &   &   & X &   &   &   &   & X &   & X \\ 
1828$+$487 &   &   &   &   &   &   &   &   &   & X &   &   \\ 
1830$+$285 &   &   &   &   & X &   &   &   & X &   &   & X \\ 
1845$+$797 &   &   &   &   &   &   &   &   &   &   &   & X \\ 
								
\\								
								
1909$+$048 &   &   &   &   &   &   &   & X & X &   &   &   \\ 
1928$+$738 &   &   &   &   & X &   &   &   &   & X &   & X \\ 
1943$+$228 &   &   &   &   & X &   &   &   & X & X &   & X \\ 
1947$+$079 &   &   &   &   &   &   &   &   &   & X &   & X \\ 
								
\\								
								
2005$+$403 &   &   &   &   & X &   &   &   & X &   &   & X \\ 
2008$-$068 &   &   &   &   &   &   &   & X &   & X &   &   \\ 
2032$+$107 &   &   &   &   & X &   &   &   &   & X &   & X \\ 
2037$+$511 &   &   &   &   &   &   &   &   &   & X &   &   \\ 
2047$+$098 &   &   &   &   &   &   & X &   &   & X &   & X \\ 
2059$+$034 &   &   &   &   &   &   &   &   &   & X &   & X \\ 
								
\\								
								
2113$+$293 &   &   &   &   &   &   & X &   &   &   &   &   \\ 
2121$+$053 &   &   &   &   & X &   &   &   & X &   &   &   \\ 
2134$+$004 &   &   &   &   &   &   &   &   & X & X &   &   \\ 
2155$-$152 &   &   &   &   &   &   &   &   & X & X &   & X \\ 
								
\\								
								
2200$+$420 &   &   &   &   & X &   &   &   &   &   &   &   \\ 
2214$+$350 &   &   &   &   & X &   &   &   &   & X &   &   \\ 
2234$+$282 &   &   &   &   &   &   &   &   & X & X &   & X \\ 
2251$+$158 &   &   &   &   & X &   &   &   & X &   &   &   \\ 
2251$+$244 &   &   &   &   &   &   &   &   & X & X &   & X \\ 
								
\\								
								
2319$+$272 &   &   &   &   & X &   &   &   & X &   &   & X \\ 
2344$+$092 &   &   &   &   & X &   &   &   &   &   &   & X \\ 
2352$+$495 &   & X & X &   & X &   &   &   & X &   &   & X \\ 
 
\enddata
\tablecomments{Only sources for which at least one ARIMA model was
identified are shown, and only ARIMA models that describe at least one 
source are listed.}
\end{deluxetable}

\begin{deluxetable}{ccccc}
\tablewidth{0pc}
\tablecolumns{5}
\tablecaption{Sources for which no ARIMA Model Identified\label{tab:noarima}}
\tablehead{\multicolumn{5}{c}{2.25~GHz}}

\startdata

0836$+$710 & 0954$+$658 & 1225$+$368 & 1328$+$254 & 1438$+$385 \\
1455$+$247 & 1611$+$343 & 1742$-$289 & 2234$+$282 & 2251$+$158 \\
2352$+$495 \\

\cutinhead{8.3~GHz}

0035$+$413 & 0528$+$134 & 1328$+$254 & 1742$-$289 & 2105$+$420 \\
2209$+$081 \\

\enddata
\end{deluxetable}

We first assess the relevance of these ARIMA models to the shocked jet 
models and the analyses described earlier, focussing on
\objectname[]{BL~Lac} (\objectname[]{2200$+$420}),
\objectname[3C]{3C~279} (\objectname[]{1253$-$055}), and
\objectname[Ohio]{OT~081} (\objectname[]{1749$+$096}).  We shall use
only the 8.3~GHz results in this comparison, both because it is a
frequency common to the University of Michigan Radio Astronomy
monitoring program and this monitoring program and because the
2.25~GHz light curves are affected more strongly by scintillation
(\S\ref{sec:sf}).

Only a single model is identified for both \objectname[]{1253$-$055}
(1,2,1) and \objectname[]{2200$+$420} (0,1,1).  For
\objectname[]{1749$+$096} two models are identified, (1,1,1)
and~(2,2,2).  \cite{haa91} found the opacity within
\objectname[]{2200$+$420} to be less that that of the other two
sources.  We would therefore expect that the effect of shocks would
not persist as long.  \cite{haa89a} show this explicitly for a
two-shock simulation---in their models, at higher frequencies (lower
opacities) shocks are more distinct.  Thus, at a given time the
observed emission from a low opacity flow is less likely to depend
upon its previous values.  Exactly that result is observed for the
ARIMA models for these three sources.  The low order of the
\objectname[]{2200$+$420} model and lack of any MA component to the
model indicate that the flux density at any given time is not strongly
dependent on its values at previous times.

Like \cite{haa92}, we conclude that shocked jet models are widely
applicable.  A key point regarding Tables~\ref{tab:arima2}
and~\ref{tab:arima8} is that most sources can be described by a
relatively small number of models.  We have constrained $p$, $d$,
and~$q \le 2$.  Thus, for each light curve there are 26 possible
models (not counting $p = q = d = 0$) by which it might be described.
Yet only 12 models are identified for all sources, and 4
models describe most sources: (0,1,1), (2,1,2), (0,2,1), and~(2,2,2).
Of these 4 most popular models, (0,1,1) describes
\objectname[]{2200$+$420} and (2,2,2) describes
\objectname[]{1749$+$096} further increasing our confidence in the
widespread applicability of shocked jet models.

A similar situation occurs at~2.25~GHz.  Only 12 models are
identified, and most sources are described by 4 models: (0,1,1),
(1,1,1), (0,2,1), and~(2,2,2).  That nearly the same set of models can
identify most sources at both frequencies is an indication that the
applicability of the shocked jet model probably extends at least down
to~2.25~GHz, even though the effects of interstellar scintillation are
also becoming important at this frequency (\S\ref{sec:sf}).

For most of the sources the light curve can be identified with a $d =
1$ or $d = 2$ model.  The use of differencing (eqn.~\ref{eqn:diff})
indicates that the light curves are likely to be non-stationary,
verifying our earlier statements that many of these light curves
appear non-stationary by virtue of not having a well-defined mean.

We have checked whether different classes of sources are over- or
under-represented in any model.  In general we find no discrepancies
between the population of BL~Lacs described by each model.  Of the
total number of sources monitored, 13.4\% are BL~Lacs.  Within the
uncertainties, all models describe roughly the same fraction of
BL~Lacs.  Again, like \cite{haa92} we conclude that a similar emission 
process occurs within both BL~Lacs and quasars.  That BL~Lac objects
are identified by a variety of models may indicate other differences
within the context of the shocked jet model besides opacity variations 
as mentioned above, e.g., differences in magnetic field strength or
jet viewing angle.

Certain sources often have a large number of models identified for
them.  These sources are those monitored for the shortest duration and
include \objectname[]{0123$+$257}, \objectname[]{0130$-$171},
\objectname[]{1250$+$568}, \objectname[]{1525$+$314},
\objectname[]{1547$+$508}, and \objectname[]{2008$-$068}.  The model
identification procedure depends upon identifying when an ACF or
partial autocorrelation function has effectively dropped to zero.  The 
uncertainty in determining the value of either function at any given
lag scales as $N^{-1/2}$.  Hence, short-duration light curves are
likely to be identified with more than one kind of ARIMA model.

Two sources deserve special mention because they are \emph{not}
quasars or BL~Lac objects.  The source \objectname[]{1742$-$289}
(\objectname[]{Sgr~A${}^*$}) is the compact radio source in the
Galactic center.  At both frequencies, no model is identified for
this source.  While \objectname[]{1742$-$289} is likely to be the
radio counterpart to a supermassive black hole, it has also been long
recognized that this radio source has a much lower luminosity than
those at the cores of active galaxies.  The lack of any identified
models for \objectname[]{1742$-$289} is consistent with its emission
process being qualitatively different than those of active galaxies.

The source \objectname[]{1909$+$048} (\objectname[]{SS~433}) is a
microquasar-like source, but the compact object in it is likely to be
neutron star rather than a black hole.  At both frequencies the models
identified for this source are (2,1,0) and~(2,1,2).  That the
model~(2,1,2) is a popular model at~8.3~GHz is an indication that
certain features of the jet emission from \objectname[]{1909$+$048}
are common to radio-loud quasars and BL~Lacs as well.

Finally, we address whether any of the sources for which no ARIMA
models are identified (Table~\ref{tab:noarima}) are notable in any
manner.  Of the sources for which a model could not be identified
at~8.3~GHz, none of them are BL~Lacs, but otherwise there appears to
be no other notable aspect.  However, examination of their light
curves show all of them to be featureless with little indication of
variability.  These light curves are exactly the kind that would have
been excluded by our requirement that $p$, $d$, and~$q$ could not all
be zero, i.e., these light curves are consistent with being dominated
purely by white noise.  Within the context of the shocked jet model,
these sources must be in a period of extended quiescence in which no
shocks occur or numerous small shocks occur continually so that the
average flux density level remains essentially unchanged.

One potentially puzzling aspect of these identifications is that ARIMA
models are identified for \objectname[]{1328$+$307} but not for
\objectname[]{1328$+$254}, even though their light curves are
superficially similar.  Close examination of the variability measures
we have employed for these two sources shows, however, that
\objectname[]{1328$+$307} is slightly more variable; this slight
difference is apparent in examination of the levels of the structure
functions on long time scales (Figure~\ref{fig:sf}).  This additional
variability is not due solely to outlier points in the light curves.
We have edited out the most visually apparent outliers from the light
curve of \objectname[]{1328$+$307}, and it remains more variable than
\objectname[]{1328$+$254}.  We conclude that the difference in ARIMA
models reflects a difference, albeit at a low level, in the
variability (i.e., the presence of shocks in the shocked jet model) in
these two sources.

We conclude that the sources' emission mechanism(s) is likely to be
broadband because the same ARIMA model is identified at both
frequencies for most sources.  The extent to which the sources have a
common emission mechanism is less clear.  Only a small number of
models is required to describe most sources, but no model describes
more than about 50\% of all sources.  Only about 20\% of the sources
are not identified with one of the three most successful models:
(0,1,1), (0,2,1), and (2,2,2).  We regard this as suggestive evidence
that there are a limited number of emission mechanisms that power
these sources.  It may also be the case that multiple mechanisms are
operative in the same source, with differences between sources due to
which process is dominant.  However, our analysis has not yielded what
we regard as unambiguous evidence of a single process powering all
sources.

\subsection{Structure Functions (Figure~\ref{fig:sf})}\label{sec:sf}

The structure function is useful for removing slow trends in a time
series and identifying short time scale fluctuations superposed upon
the slow trends.  The first-order temporal structure function of the
flux density~$F$ is
\begin{equation}
D_F(t, \tau) = \langle[F(t) - F(t+\tau)]^2\rangle
\label{eqn:sf}
\end{equation}
(see \citealt{sch85} for a description of higher order structure
functions).  The first-order structure function removes any linear
dependences and reflects the presence of higher-order polynomial terms
and stationary processes in the time series, and the slope of the
logarithmic structure function is a measure of the temporal structure
in the light curve.  For a stationary, zero-mean time series with a
variance of $\sigma_F^2$, the structure function is related to the
correlation function~$\rho_F(\tau)$ by
\begin{equation}
D_F(\tau) = 2\sigma_F^2[1 - \rho_F(\tau)]^2.
\label{eqn:sfrho}
\end{equation}
As we discuss above (\S\ref{sec:arima}), the light curves of many of
these sources are unlikely to be stationary.  For that reason we have
employed the definition of the structure function given in
equation~(\ref{eqn:sf}).  A ``break'' in the structure function can be
used to identify the characteristic time scale for a process
contributing to the time series to saturate (e.g., $\rho[\tau] \to
0$).

Figure~\ref{fig:sf} shows the resulting structure functions.  Part of
the value of the structure function is in identifying short time scale
fluctuations.  For this reason we have used the \emph{unsmoothed}
light curves in forming the structure functions.  In constructing
these structure functions, we have not used any pairs of data for
which one datum was before the receiver change and the other after the
receiver change.  We also have normalized the structure functions by
the variance of the time series.  The uncertainties displayed on the
structure functions are calculated as \citep{sch85}
\begin{equation}
\sigma_D^2(\tau) \approx \frac{8\sigma_F^2}{N_D(\tau)}D_F(\tau)
\label{eqn:sfrms}
\end{equation}
where $\sigma_F^2$ is the variance of the time series and $N_D(\tau)$
is the number of data pairs used to form the structure function at
lag~$\tau$.

\citeauthor{fiedleretal87} found that, for a subsample of the sources
presented here, the typical logarithmic slope of the structure
function (calculated using equation~[\ref{eqn:sfrho}]) was 1--1.5 for
lags less than 20~days.  Such a slope is consistent with interstellar
scattering resulting from a scattering medium distributed along the
line of sight (slope $\approx 1$) as opposed to scattering distributed
in a screen (slope $\approx 2$; \citealt{sch85}; \citealt{bnr86}).
Uncorrelated variations would have a logarithmic slope of~0.

We have determined the logarithmic slope for lags ranging between~2
and~32~days.  We have used a maximum lag of either 16 or~32~days
(i.e., bracketing the maximum lag considered in
\citeauthor{fiedleretal87}) and minimum lags of~2 and~4~days.
Because of the day-to-day variability introduced by instrumental
effects, the value of the structure function may contain a
contribution from the instrumental variability in addition to any
scattering-induced or intrinsic variability.  We find that the
logarithmic slope for our larger sample is approximately 0.3, with the
choice of maximum and minimum lags making little difference.  The
maximum logarithmic slope for any source is 1.3.

While consistent with a scattering-induced variability on short time
scales, the above argument does not rule out an intrinsic contribution 
to the sources' short-term variability.  In an effort to discriminate
between an extrinsic and intrinsic origin of the short-term
variability, we have compared the structure function slopes to the
Galactic latitudes and spectral indices of the sources.

\begin{figure}[tbh]
 \begin{center}
 \mbox{\psfig{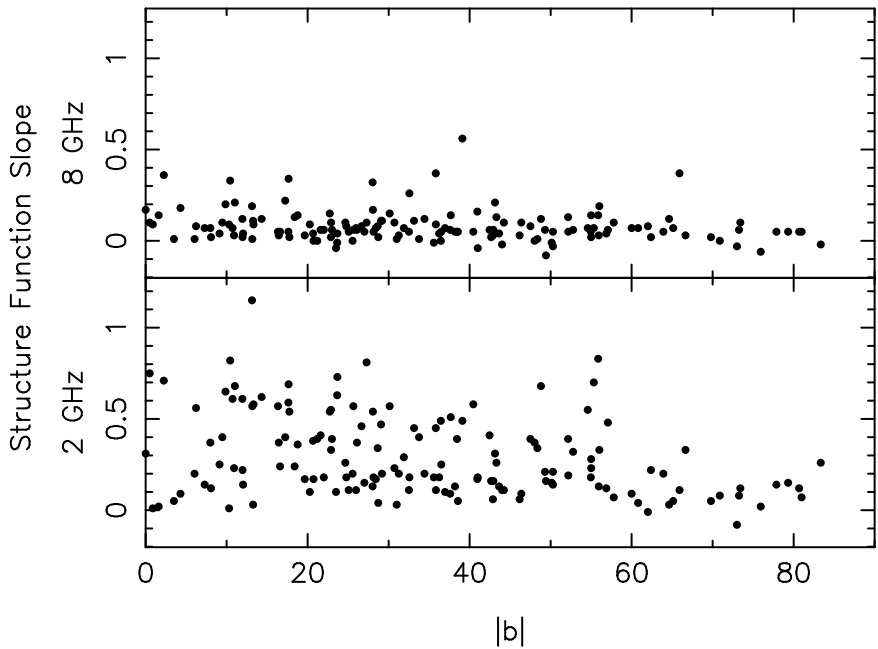}} 
 \end{center}
\vspace{-0.5cm}
\caption[]{Structure function slopes as a function of
(absolute) Galactic latitude.  Shown are the logarithmic structure
function slopes computed for lags between~2 and~32 days.  
\textit{Top}: X-band (8~GHz); 
\textit{Bottom}: S-band (2~GHz).}
\label{fig:slopeb}
\end{figure}

\begin{figure}[tbh]
 \begin{center}
 \mbox{\psfig{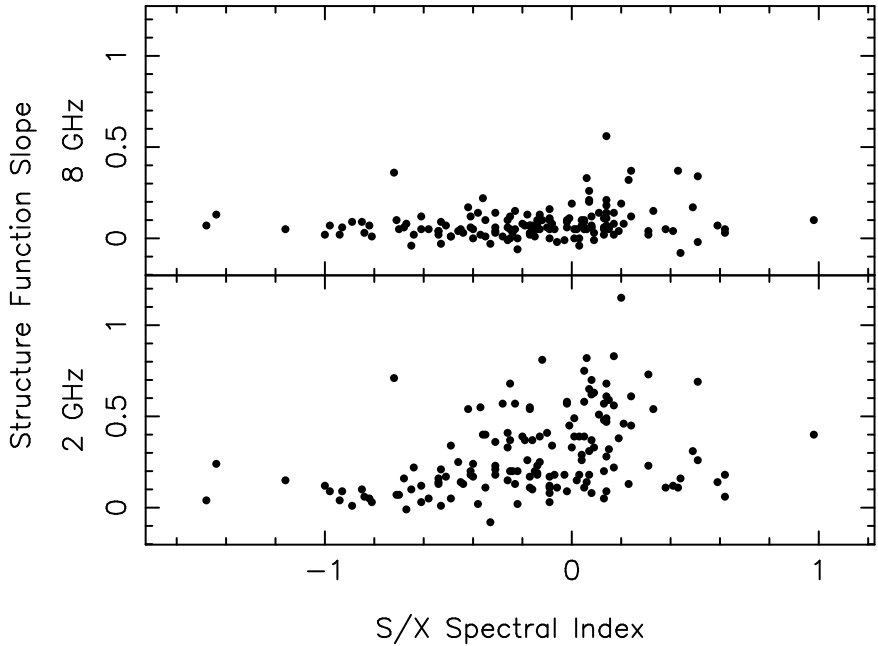}}
 \end{center}
\vspace{-0.5cm}
\caption[]{Structure function slopes as a function of
S/X spectral index.  As for
Figure~\ref{fig:slopeb}.}
\label{fig:slopesi}
\end{figure}

Figure~\ref{fig:slopeb} shows the structure function slopes for all
sources as a function of (the absolute value of) Galactic latitude,
and Figure~\ref{fig:slopesi} shows the structure function slopes for
all sources as a function of the S/X spectral index.  These figures
were constructed from the structure function slopes computed using a
2-day minimum lag and 32-day maximum lag.  Our expectation is that
there should be a latitude dependence if scattering is at least
partially responsible for these short-term variations, and a
spectral-index dependence if intrinsic variations are at least
partially responsible.

There is a strong correlation between the structure function slopes
at~2.25~GHz and Galactic latitude.  The correlation coefficient is $r
= -0.34$; this strong correlation increases if we bin the data.  We
can exclude, at better than the 99\% confidence level, the hypothesis
that the structure function slopes at~2.25~GHz and the sources' Galactic
latitudes are uncorrelated.  No significant correlation exists between
the structure function slopes at~8.3~GHz and Galactic latitude.  A
strong correlation also exists between the structure function slopes
at~2.25~GHz and the S/X spectral index, but not at~8.3~GHz.  (We have
also verified that there is no correlation between the sources'
Galactic latitude and spectral indices which might conspire to produce
the correlations with structure function slopes.)

The strong correlation is essentially unchanged if we remove
\objectname[]{1741$-$038} ($b = 13\arcdeg$) from the analysis.  This
source underwent an ESE in 1992, and its inclusion might have been
thought to bias the correlation.

We conclude that scattering plays a significant role in causing the
short-term ($\sim 10$~d) variability of these sources or at least
those at low latitude.  We cannot rule out intrinsic variability also
contributing to the short-term variations.  If intrinsic variations
represent a significant fraction of the short-term variability,
however, given that flat-spectrum sources tend to be more variable at
higher frequencies, the lack of a stronger correlation between the S/X
spectral indices and the structure function slopes at~8.3~GHz is
puzzling.

Figure~\ref{fig:slopeb} also shows that we are in general agreement
with \citeauthor{fiedleretal87} regarding the distribution of
scattering toward these sources.  At low latitudes, the logarithmic
slope increases toward unity, as expected if scattering is better
described as by volume than a single screen.  The large scatter at low
latitudes is due probably to the inhomogeneous distribution of
scattering.  Our lower value for the average logarithmic slope can be
attributed to the larger number of high latitude sources for which
scattering is likely to be less important.

\subsection{Wavelet Transform Identification of Extreme Scattering Events}\label{sec:wavelet}

One of the primary motivations for conducting this series of
observations was to identify additional extreme scattering events
\citep{fdjh87}.  Extreme scattering events identified thus far from
this data set are shown in \cite{fdjws94} and \cite{cfl98}.  However,
all previous ESEs have been identified visually.  Here we attempt a
systematic search for ESEs using a wavelet transform.  Because of the
strong wavelength dependence of ESEs, we focus on the 2.25~GHz light
curves for this analysis.

A wavelet transform of a discretely sampled time series~$F(t_i)$ is \cite[e.g.,][]{s89}
\begin{equation}
W_F(s, l) = \sum_i \psi(s, l; t_i) F(t_i).
\label{eqn:wavelet}
\end{equation}
The wavelet transform kernel~$\psi(s,l)$ depends upon both a ``scale''
size~$s$ and a location~$l$, in contrast to a Fourier transform
kernel which depends upon only the frequency or scale size.  As a
consequence, wavelet transforms are useful in detecting features that
are localized in time, including transient events.

In order to account for gaps in the time series, we followed a
procedure suggested by \cite{s89} for the autocorrelation function of
a time series with missing data.  We calculated $W_F(s, l)$ and
$W_U(s, l)$, where $U(t_i)$ is a time series with the same sampling as
$F(t_i)$ but with unit flux density.  The analyzed quantity was
$(W_F/W_U)$.  The wavelet coefficients $W_U$ account for the uneven
sampling of the time series; forming the ratio decreases the magnitude
of coefficients~$W_F$ which occur largely because of the uneven
sampling.

For the purposes of identifying ESEs, we compared two different wavelet
kernels---the Haar and the mexican hat (Marr) bases.  The former is
a unit increase followed by a unit decrease and resembles the ingress
(or the negative of the egress) of a (symmetric) \hbox{ESE}.  The
latter is the second derivative of a gaussian and closely resembles
the negative of a (symmetric) \hbox{ESE}.  We found that the mexican
hat basis produced a larger response to ESEs already identified, and
we shall use it in our analysis.

The wavelet coefficients themselves are the sum of a number of flux
densities.  The individual flux density measurements are nearly
gaussian distributed (\citeauthor{fiedleretal87}), and, by the central
limit theorem, their sum will be even more nearly gaussian
distributed.  We exploit the gaussian distribution of the wavelet
coefficients to search for ESEs in the following manner: For each
scale~$s$ we have determined the mean and variance, $\sigma_W^2$,
averaged over all locations~$l$.  Wavelet coefficients having less
than a 99\% probability of occurring by chance, i.e., those that
deviate by more than $2.57\sigma_W$ from the mean, were considered to
be potential ESEs.

The known ESEs have durations of roughly a few weeks to a few months.
This typical duration is almost certainly a selection effect, as the
shortest ESE that can be identified depends upon the sampling
frequency and the longest ESE depends upon the duration of the light
curve.  With our typical sampling of one flux density measurement
every two days, we will be unable to identify reliably ESEs with a
duration of 8~days or less.  For this reason we both smoothed the
light curves by an 8-day boxcar (this also reduces the effects of the
day-to-day variability) and considered only wavelet coefficients for
scales 16~days or longer.  The maximum duration of an ESE that we
could detect is limited by the length of the monitoring program.  Our
ESE detection procedure also requires that there be a reasonable
number of wavelet coefficients to form the mean and variance.  For
sources observed during 1988--1994, we restricted the largest scale to
be $s \le 128$~days; for the sources whose observations commenced
prior to~1988, we allowed $s \le 1024$~days.

The wavelet coefficients identified from this procedure as being
significant comprise a set of candidate ESEs.  However, instrumental
changes or malfunctions could also result in a significant wavelet
coefficient.  In many cases, significant wavelet coefficients were
found on the same or adjoining days for sources separated by tens of
degrees on the sky.  If two or more sources showed significant wavelet
coefficients occurring within 4~days (i.e., 2 samples) of each other,
we eliminated these wavelet coefficients from the set of candidate
ESEs.  Because an outburst from a source could also produce a
significant coefficient, we examined visually the light curves in the
vicinity of the significant coefficients.  We eliminated wavelet
coefficients that did not show a decrease in flux density, as would be
expected from an \hbox{ESE}.

Table~\ref{tab:ese} lists the location of probable ESEs that survived
this culling.  Figure~\ref{fig:ese} shows the light curves in the
vicinity of these ESEs.

\begin{deluxetable}{lcc}
\tablewidth{0pc}
\tablecaption{Wavelet-Identified ESEs in the GBI Monitoring Program\label{tab:ese}}
\tablehead{
	\colhead{Name} & \colhead{Epoch} & \colhead{Significance} \\
	               &                 & \colhead{($\sigma_W$)}}

\startdata

\objectname[]{0133$+$476} & 1990.2 & 8.8 \\
\objectname[]{0201$+$113} & 1991.3 & 3.5 \\
\objectname[]{0202$+$319} & 1989.5 & 6.7 \\
\objectname[]{0300$+$470} & 1988.3 & 6.3 \\
\objectname[]{0528$+$134}\tablenotemark{a} & 1991.0 & 2.9 \\
	   & 1993.9 & 9.6 \\
\\

\objectname[]{0952$+$179}\tablenotemark{a} & 1992.6 & 7.4 \\
           & 1993.6 & 9.6 \\
\objectname[]{0954$+$658} & 1981.1 & 15.9 \\
\objectname[]{1438$+$385} & 1993.5 & 3.4 \\
\objectname[]{1502$+$106} & 1979.5 & 6.2 \\
\objectname[]{1756$+$237} & 1993.7 & 2.8 \\

\\

\objectname[]{2251$+$244} & 1989.9 & 3.4 \\
\objectname[]{2352+495} & 1984.6 & 3.6 \\
\enddata

\tablenotetext{a}{Multiple ESEs were identified for this source.}

\tablecomments{Column~(3), Significance, shows the number of standard
deviations that the wavelet coefficient exceeds the mean.}
\end{deluxetable}

\begin{figure}[tbh]
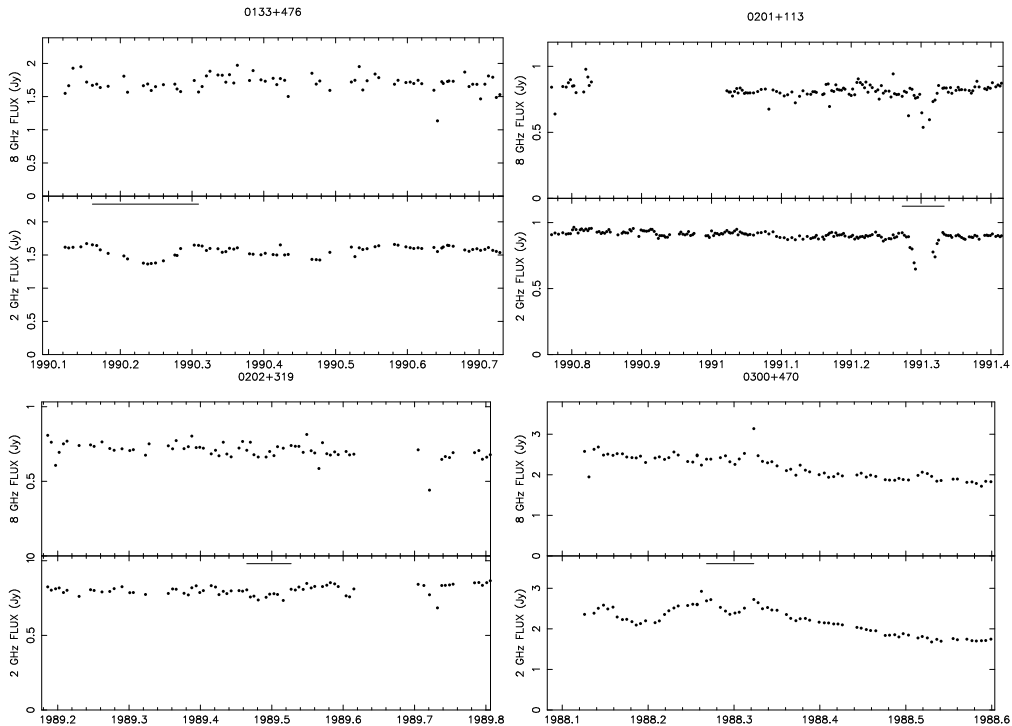

 \begin{center}
 \mbox{\psfig{file=Lazio_f7a.ps,width=0.4\textwidth,angle=-90,silent=}
       \psfig{file=Lazio_f7b.ps,width=0.4\textwidth,angle=-90,silent=}}
 \mbox{\psfig{file=Lazio_f7c.ps,width=0.4\textwidth,angle=-90,silent=}
       \psfig{file=Lazio_f7d.ps,width=0.4\textwidth,angle=-90,silent=}}
 \end{center}
\vspace{-0.5cm}
\caption[]{Light curves in the vicinity of probable extreme
scattering events identified from the wavelet analysis.  A horizontal
bar near the top of the lower panel indicates the approximate duration
of the \hbox{ESE}.}
\label{fig:ese}
\end{figure}

\setcounter{figure}{6}
\begin{figure}[tbh]
 \begin{center}
 \mbox{\psfig{file=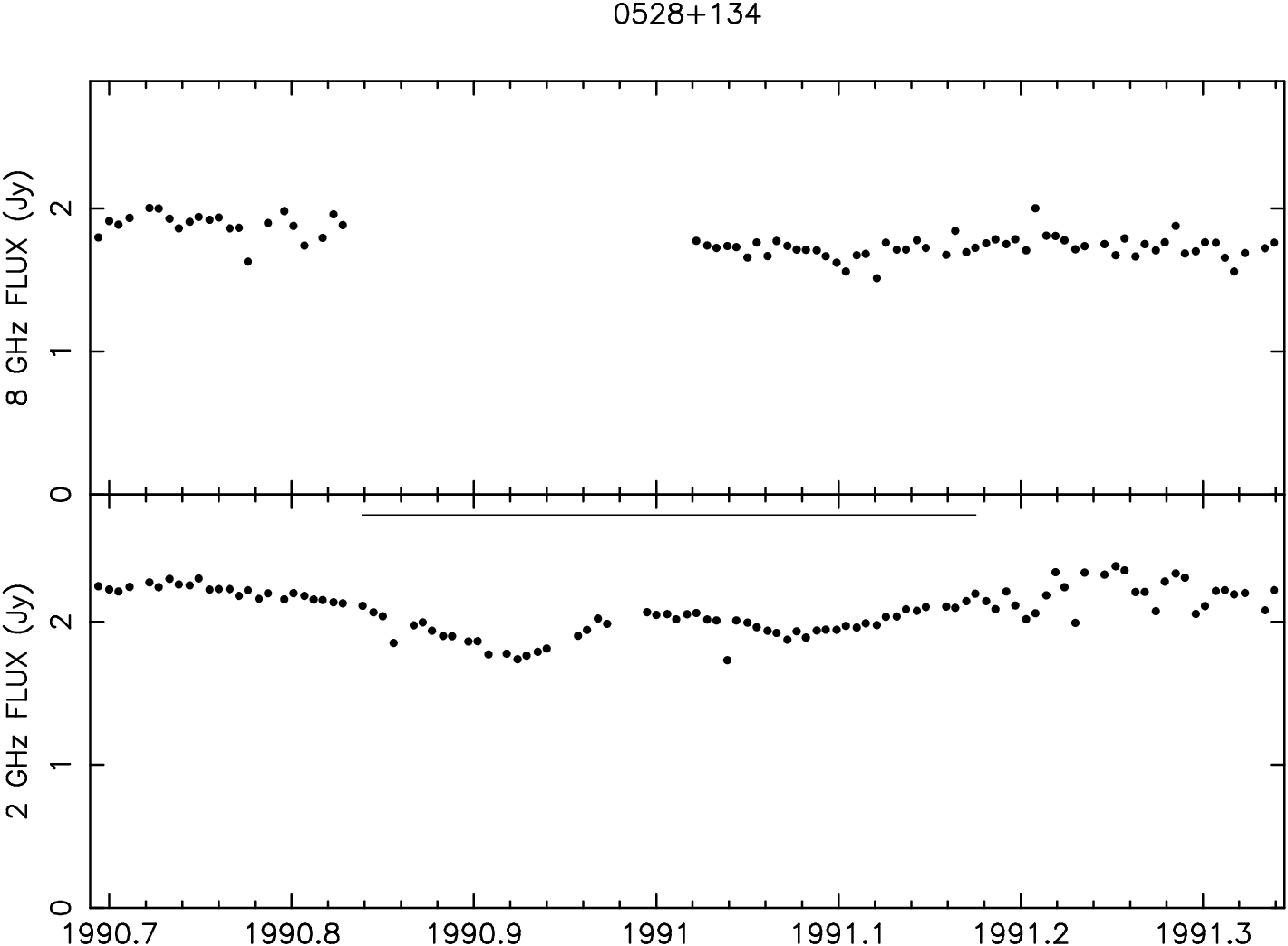,width=0.4\textwidth,angle=-90,silent=}
       \psfig{file=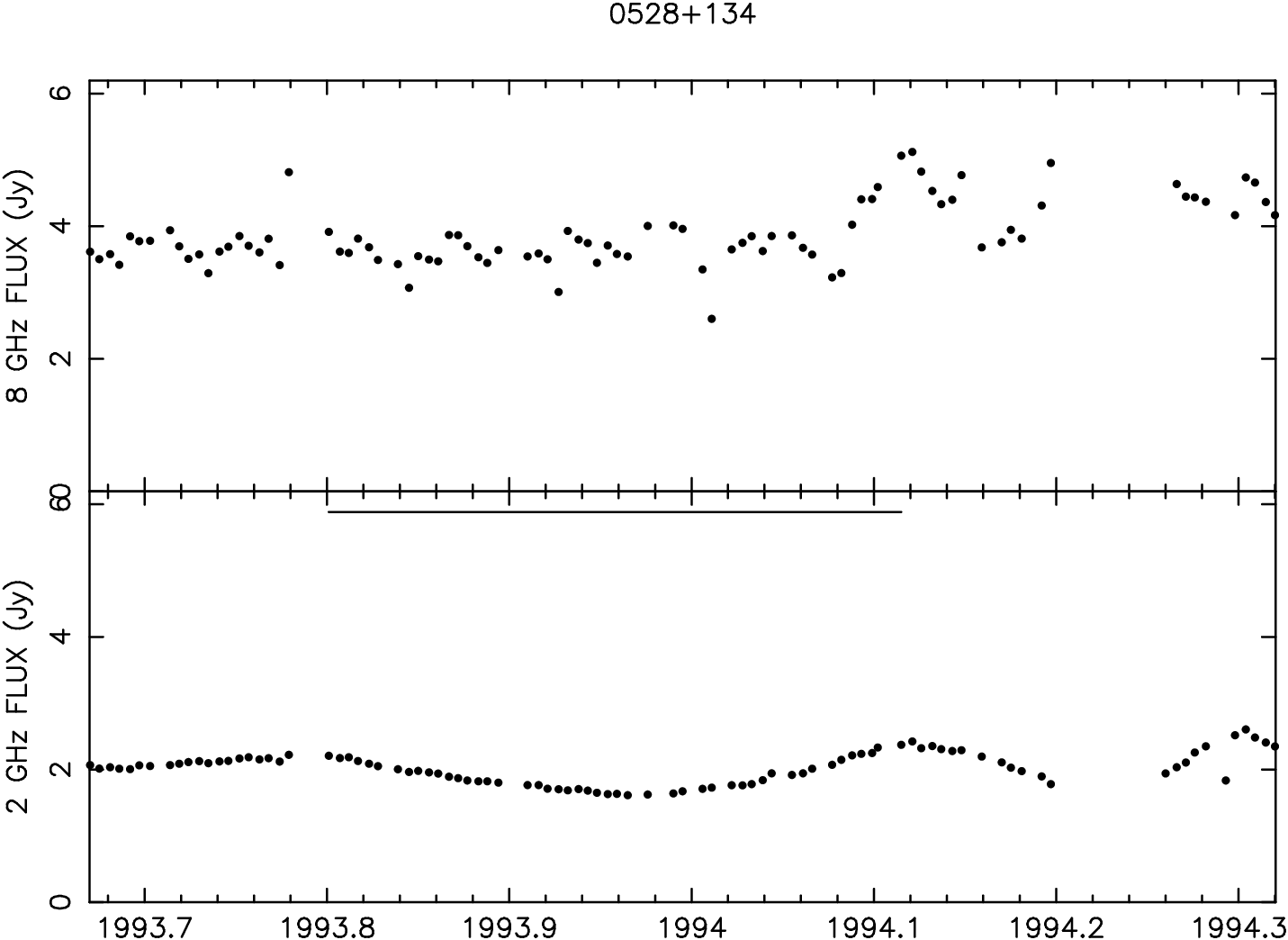,width=0.4\textwidth,angle=-90,silent=}}
 \mbox{\psfig{file=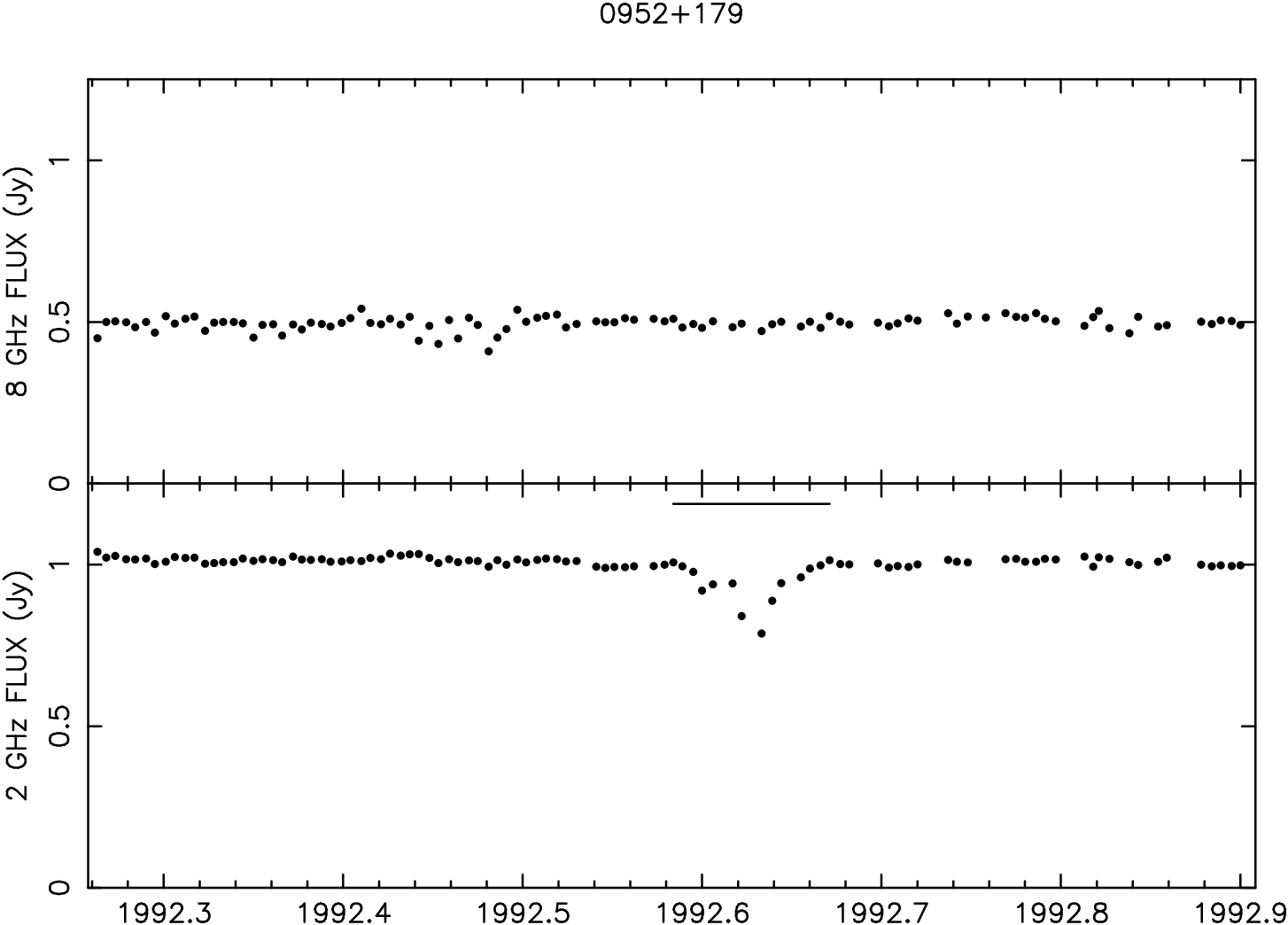,width=0.4\textwidth,angle=-90,silent=}
       \psfig{file=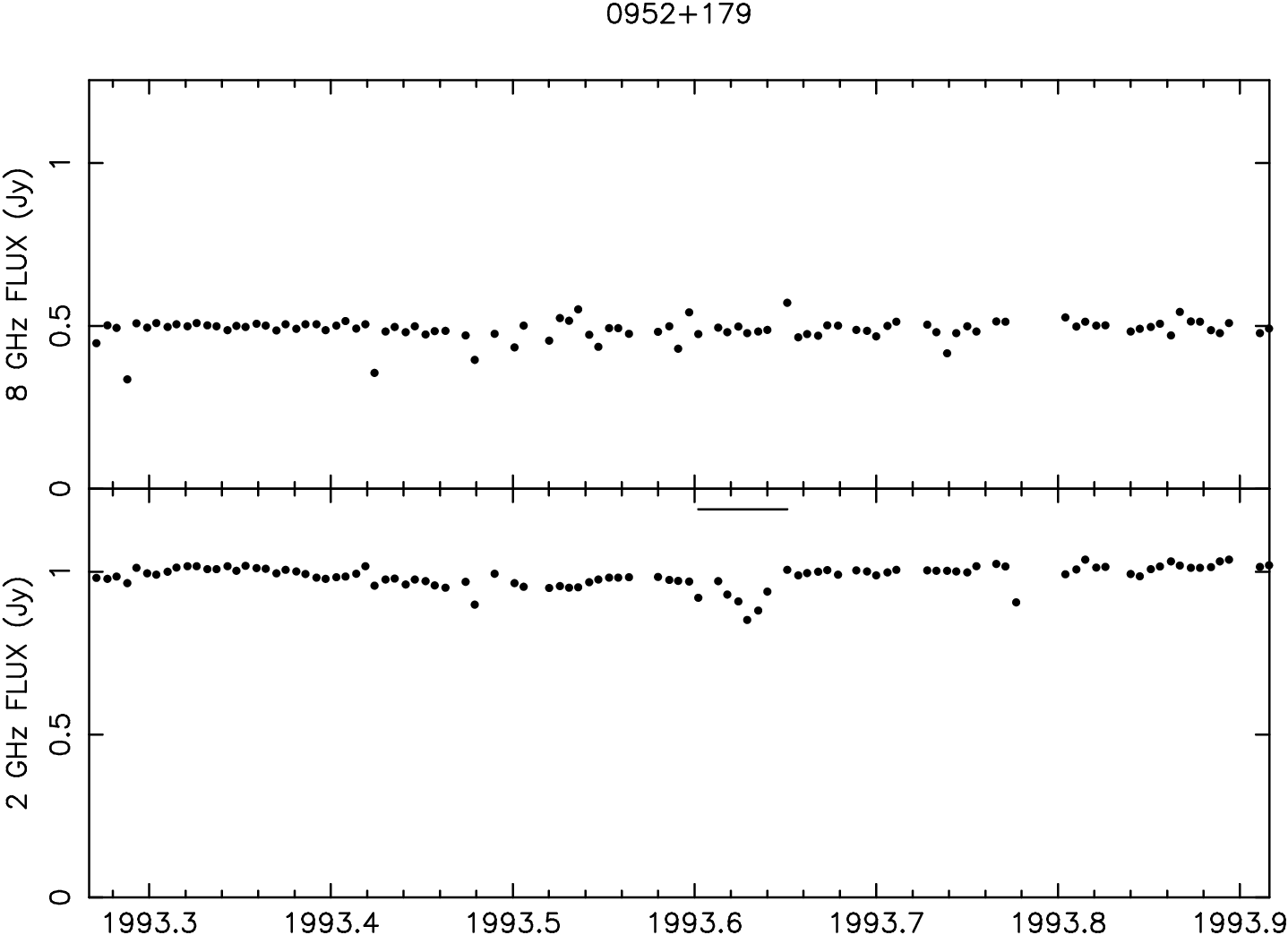,width=0.4\textwidth,angle=-90,silent=}}
 \end{center}
\vspace{-0.5cm}
\caption[]{\textit{Cont.}}
\end{figure}

\setcounter{figure}{6}
\begin{figure}[tbh]
 \begin{center}
 \mbox{\psfig{file=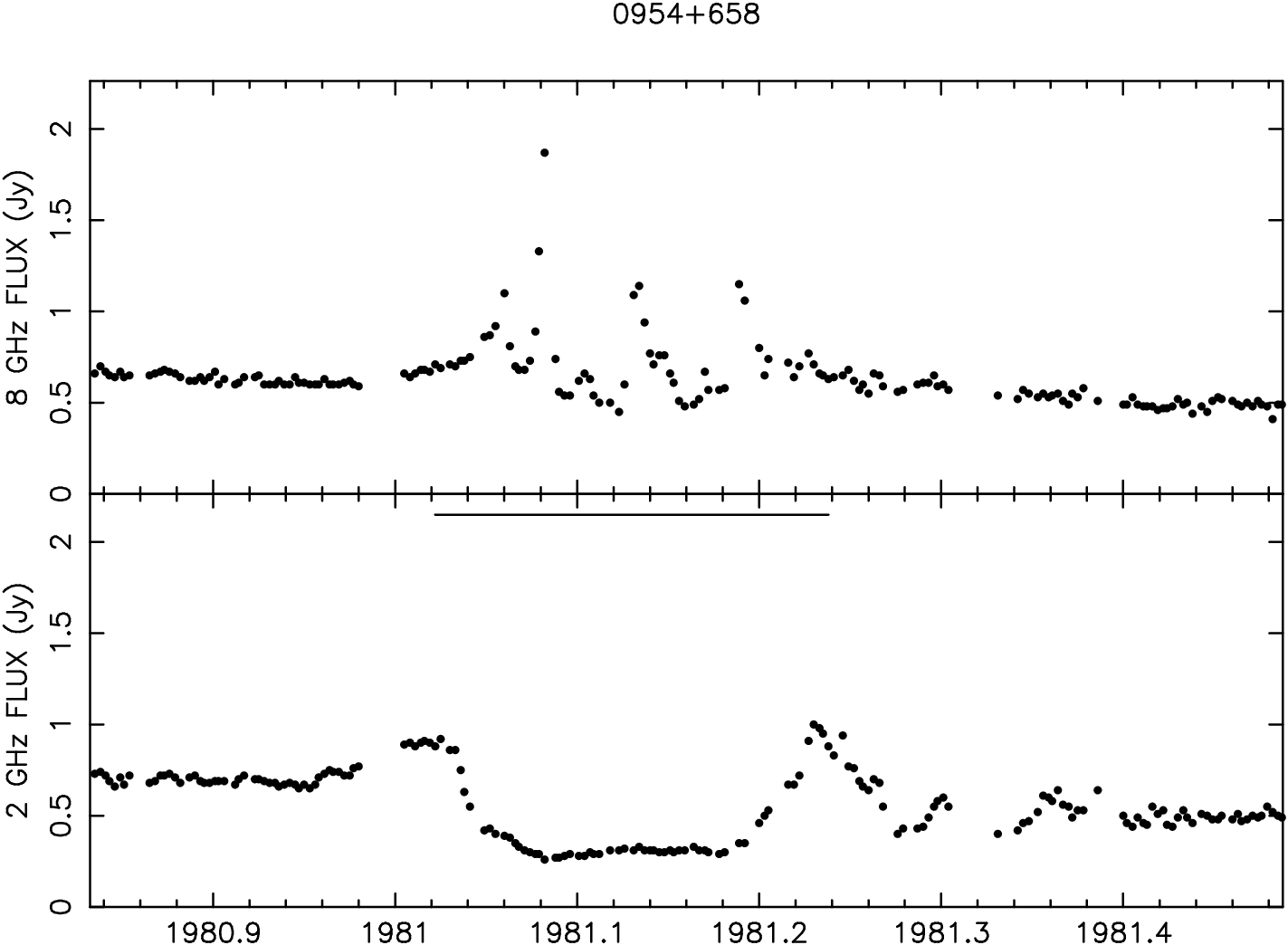,width=0.4\textwidth,angle=-90,silent=}
       \psfig{file=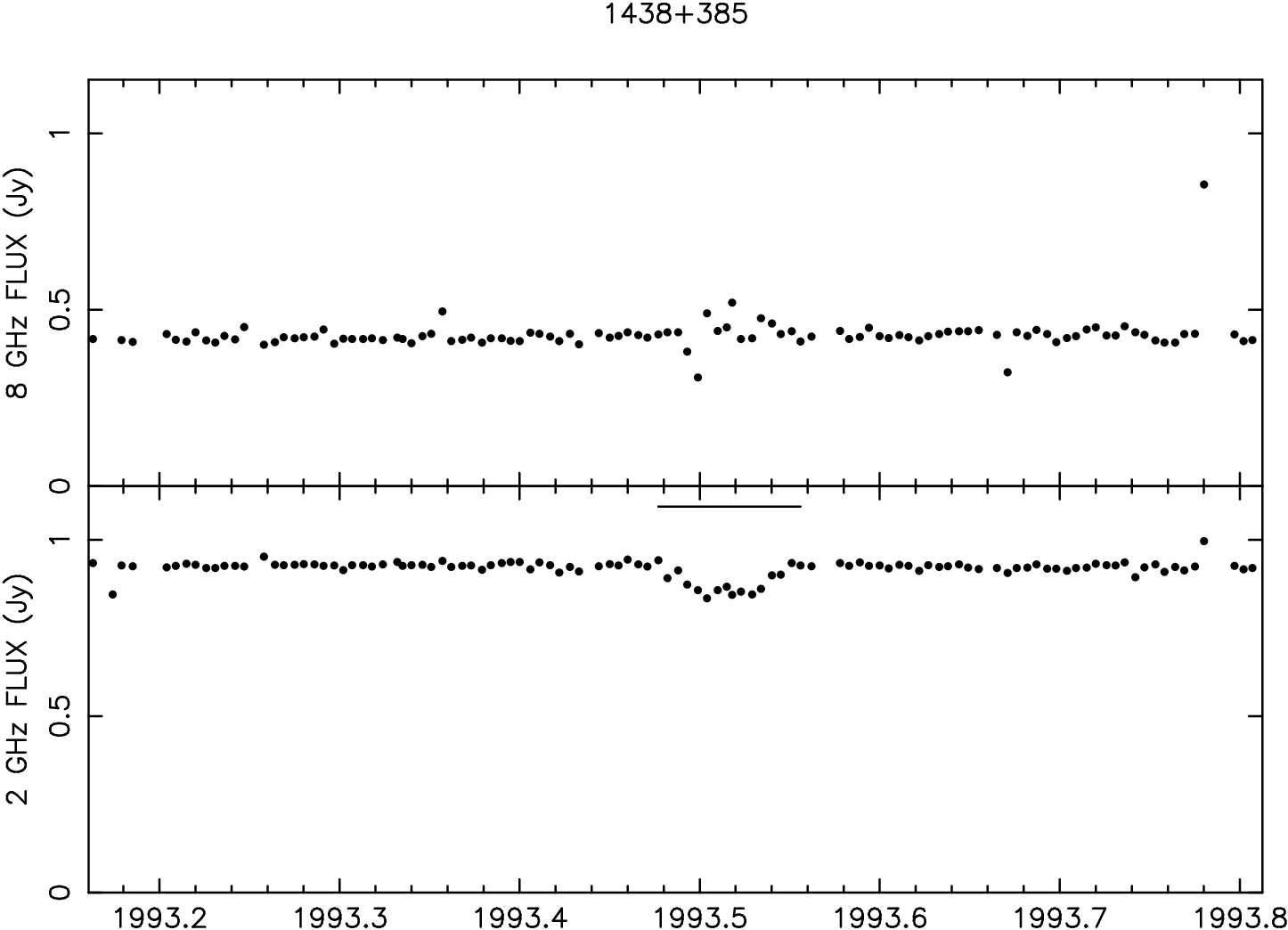,width=0.4\textwidth,angle=-90,silent=}}
 \mbox{\psfig{file=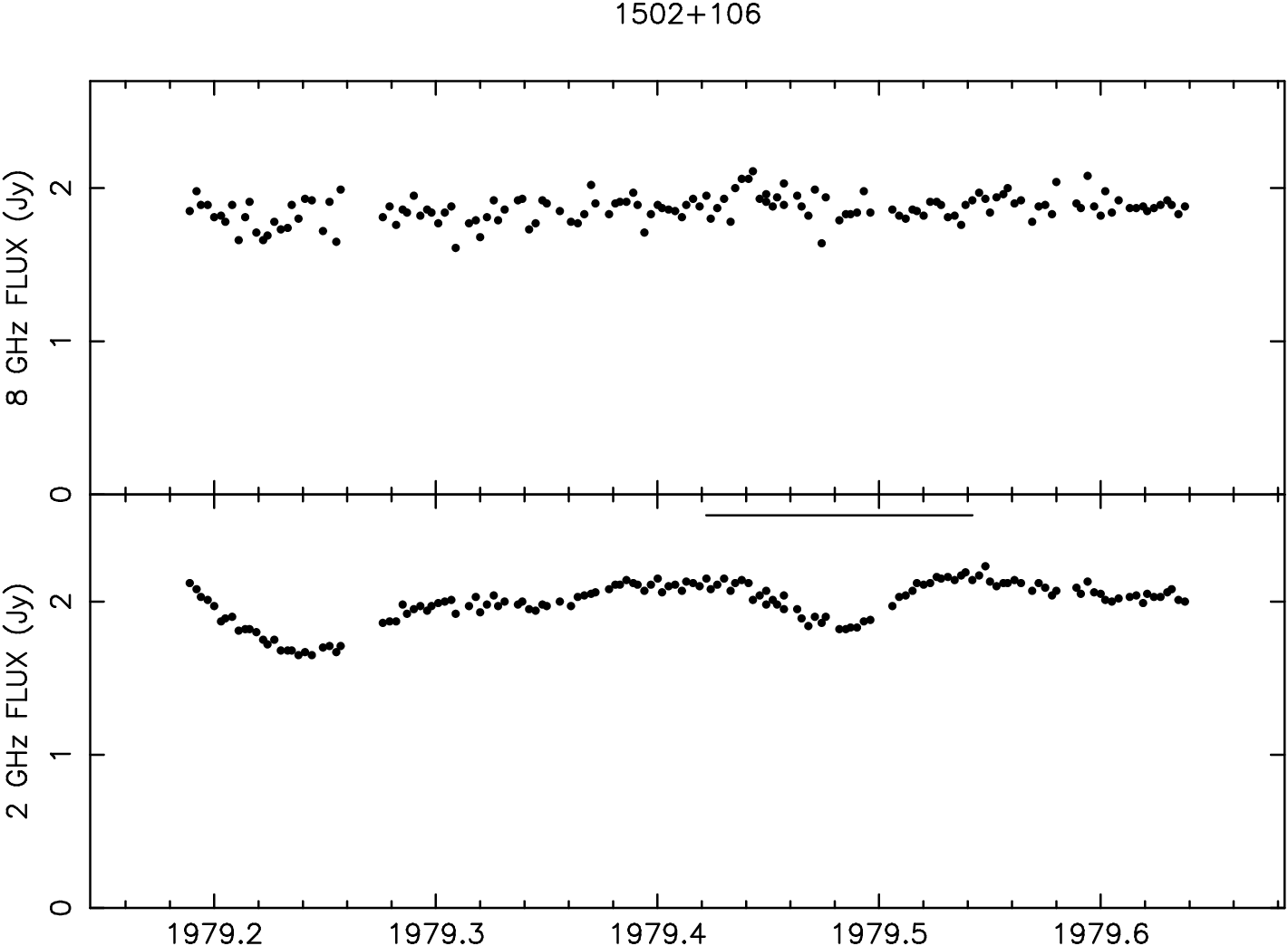,width=0.4\textwidth,angle=-90,silent=}
       \psfig{file=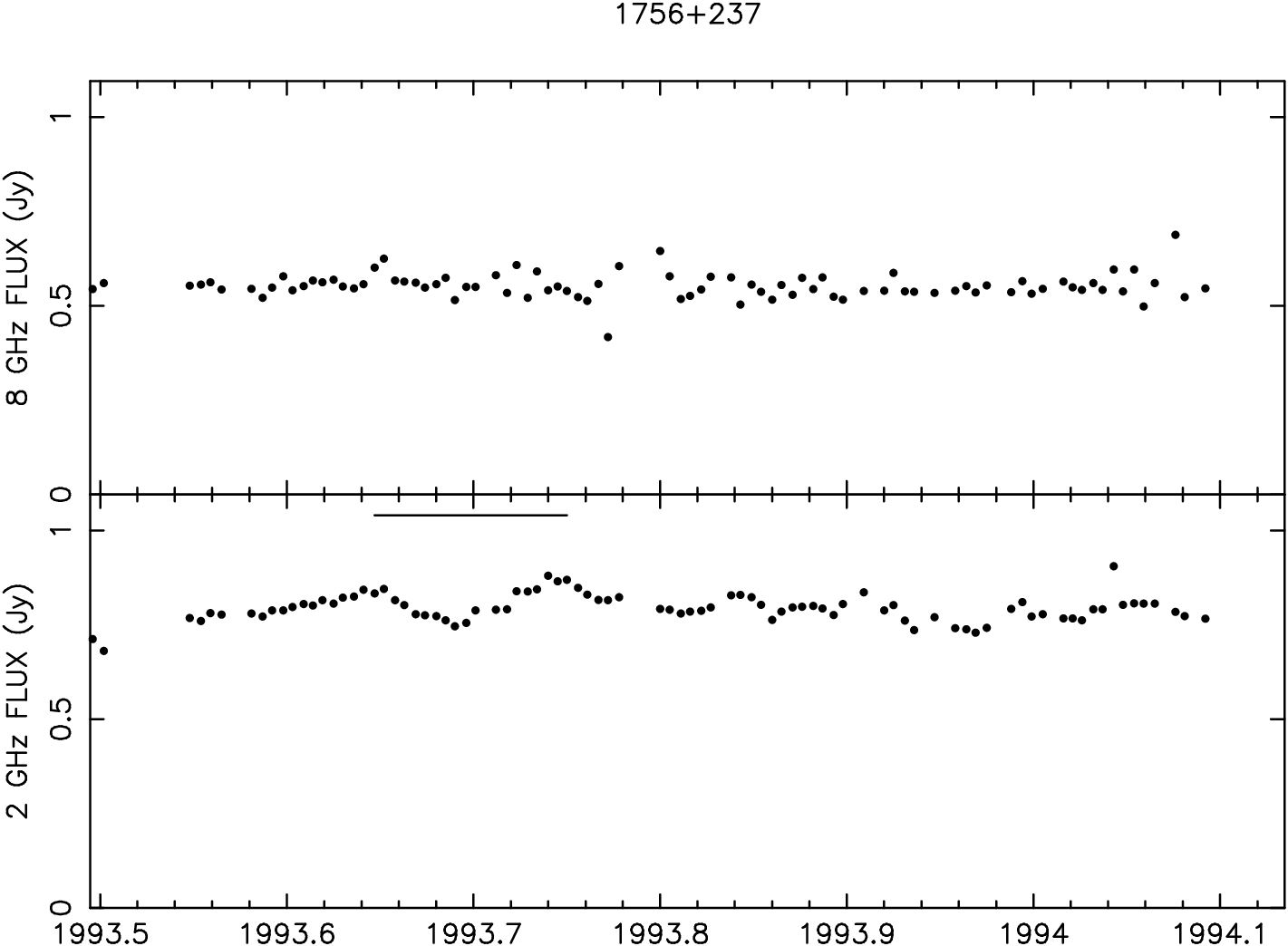,width=0.4\textwidth,angle=-90,silent=}}
 \end{center}
\vspace{-0.5cm}
\caption[]{\textit{Cont.}}
\end{figure}

\setcounter{figure}{6}
\begin{figure}[tbh]
 \begin{center}
 \mbox{\psfig{file=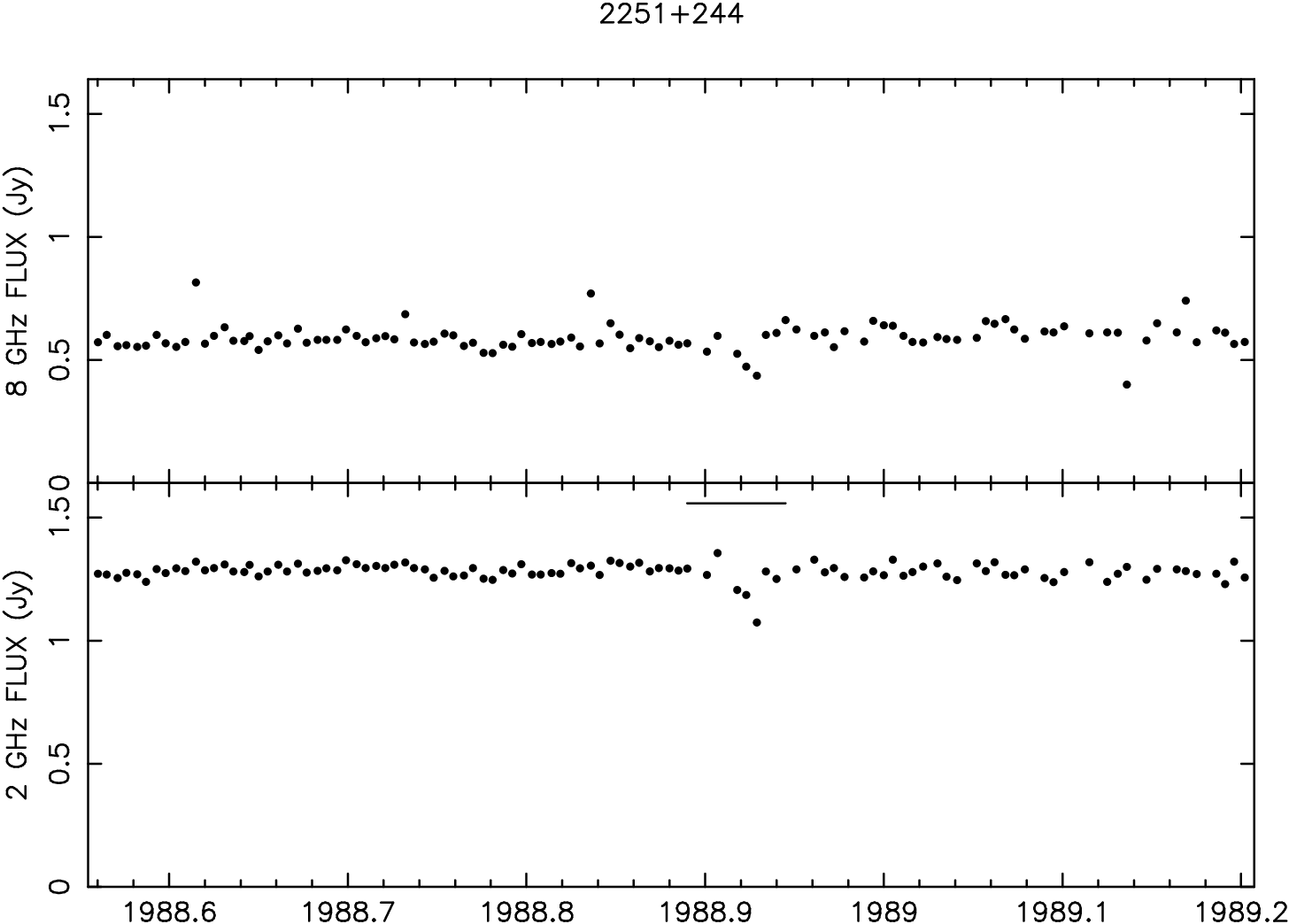,width=0.4\textwidth,angle=-90,silent=}
       \psfig{file=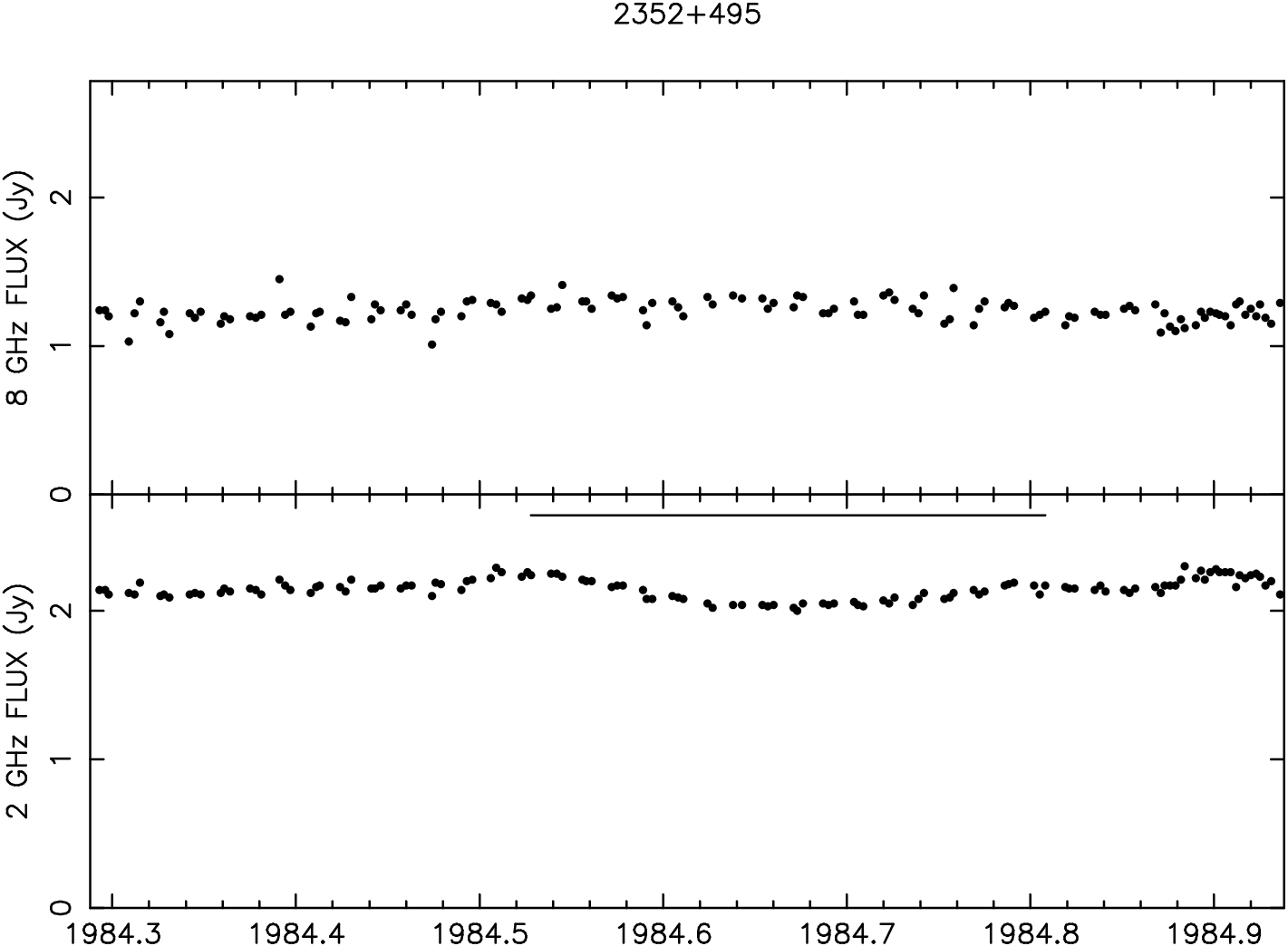,width=0.4\textwidth,angle=-90,silent=}}
 \end{center}
\vspace{-0.5cm}
\caption[]{\textit{Cont.}}
\end{figure}

\begin{figure}[tbh]
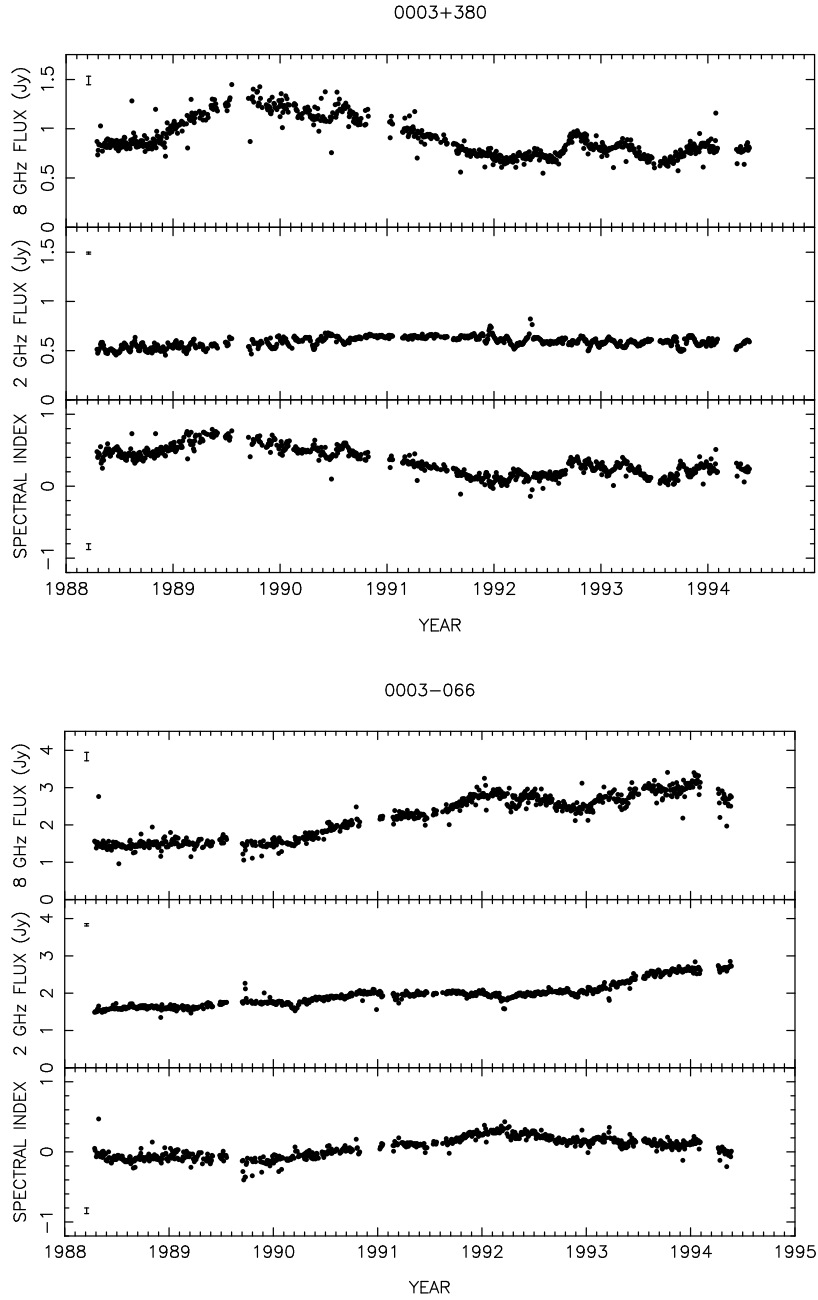

 \begin{center}
 \mbox{\psfig{file=Lazio_f8aa.ps,width=0.65\textwidth,angle=-90,silent=}}
 \linebreak\vspace{1ex}\linebreak
 \mbox{\psfig{file=Lazio_f8ab.ps,width=0.65\textwidth,angle=-90,silent=}}
 \end{center}
\caption[]{X-band (8~GHz) and S-band (2~GHz) light curves and
S/X spectral indices.  The discontinuities in flux levels at~1989.7
result from a change of observing frequency from~2.7 and~8.1~GHz
to~2.25 and~8.3~GHz.  The size of the typical measurement uncertainty
is indicated at the left of each light curve.  \emph{Only a portion of 
Figure~\ref{fig:ltcurv} is shown.  All of Figure~\ref{fig:ltcurv} and
the data used to construct it are available at
\url{http://ese.nrl.navy.mil/GBI/GBI.html}.}}
\label{fig:ltcurv}
\end{figure}

In a limited number of instances, significant wavelet coefficients
occurred at times where multiple processes may have been occurring
simultaneously.  For instance, between~1985 and~1987 the source
\objectname[]{1502$+$106} underwent an outburst
(Figure~\ref{fig:ltcurv}).  During the outburst both the 2.7
and~8.1~GHz flux densities show variations on shorter time scales.
This short time scale variability may be the result of intrinsic
variations of a newly ejected component, scintillation of a compact
component produced in the outburst \citep{r00}, and/or an ESE
\citep{fdjws94}.

Having scanned the light curves in a systematic manner for ESEs, we
can also address how many ESEs identified previously
\citep{fjwg92,fdjws94,pohletal95} can be considered significant.
Table~\ref{tab:preese} lists ESEs identified previously from the
sources included in this monitoring program.  Of the 12 ESEs
identified previously, a substantial duration of the ESE for four
sources (\objectname[]{0133$+$476}, \objectname[]{0300$+$471},
\objectname[]{0528$+$134}, \objectname[]{1749$+$096}) occurred during
a transition in the monitoring program.  Although the effect of these
ESEs can be seen in the GBI light curves, the identification of these
ESEs is almost certainly due to monitoring programs being conducted
simultaneously at other telescopes.  Thus, their absence in our list
of wavelet-identified ESEs is not surprising.  As discussed above, the
variability identified in the light curve of \objectname[]{1502$+$106}
may not be due to an \hbox{ESE}.

\begin{deluxetable}{lccl}
\tablewidth{0pc}
\tablecaption{Previously Identified ESEs from the GBI Monitoring Program\label{tab:preese}}
\tablehead{
	\colhead{Name} & \colhead{Epoch} & \colhead{Wavelet} 
	& \colhead{Comments} \\
	&              & \colhead{Identification?}}

\startdata

\objectname[]{0133$+$476} & 1988.2 &  no & occurred during GBI transition \\
\objectname[]{0300$+$471} & 1988.2 &  no & occurred during GBI transition \\
\objectname[]{0333$+$321}\tablenotemark{a}
                          & 1986.3 &  no \\
                          & 1987.9 &  no \\
\objectname[]{0528$+$134} & 1993.5 &  no & hardware malfunction, data lost \\
\objectname[]{0954$+$658} & 1981.1 & yes \\
	    	     	
\\
	    	     	
\objectname[]{1502$+$106} & 1986.0 &  no & enhanced scintillation during outburst? \\
\objectname[]{1611$+$343} & 1985.4 &  no \\
\objectname[]{1741$-$038} & 1992.5 &  no \\
\objectname[]{1749$+$096} & 1988.1 &  no & occurred during GBI transition \\
\objectname[]{1821$+$107} & 1984.2 &  no \\
	    	     	
\\
	    	     	 
\objectname[]{2352$+$495} & 1985.0 & yes & only partially identified \\

\enddata

\tablenotetext{a}{Multiple ESEs were identified for this source.}

\end{deluxetable}

Of the remaining 7 previously identified ESEs, only the ESEs toward
\objectname[]{0954+658} and~\objectname[]{2352+495} produce
significant wavelet coefficients.  Sources without significant wavelet
coefficients at the time of a visually-identified ESE are
\objectname[]{0333$+$321} (2 ESEs identified), \objectname[]{1611$+$343},
\objectname[]{1741$-$038}, and~\objectname[]{1821$+$107}.

For these latter four sources, we have examined the wavelet
coefficients in the neighborhood of their ESEs.  We find that, with
the exception of \objectname[]{1741$-$038}, there were potentially
significant wavelet coefficients that were culled by our requirement
that there be no other sources having significant wavelet coefficients
within 4~days.

In the presence of at least one instrumentally-induced variation, we
regard this requirement as necessary.  Nonetheless, we have
re-examined the light curves for ESE-like features having potentially
significant wavelet coefficients that were initially culled by this
requirement.  The epochs of these ESE-like features are listed in
Table~\ref{tab:ese_maybe}.  We consider these ESE-like features as
having an uncertain confidence level but include them here for
completeness.

\begin{deluxetable}{lc}
\tablecaption{Potential ESEs in the GBI Monitoring Program\label{tab:ese_maybe}}
\tablewidth{110pt}
\tablehead{\colhead{Name} & \colhead{Epoch}}

\startdata
0003$+$380\tablenotemark{a} 
           & 1990.3 \\
           & 1992.2 \\
0016$+$731 & 1992.5 \\
0019$+$058 & 1991.2 \\
0134$+$329 & 1988.9 \\
0256$+$075 & 1991.6 \\

\\

0333$+$321 & 1988.6 \\
0403$-$132 & 1993.0 \\
0444$+$634 & 1992.0 \\
0454$+$844 & 1993.4 \\
0528$+$134\tablenotemark{a} 
           & 1992.4 \\
           & 1994.0 \\

\\

0555$-$132 & 1993.5 \\
1438$+$385 & 1993.5 \\
1502$+$106 & 1994.0 \\
1538$+$149 & 1991.0 \\
1614$+$051 & 1992.8 \\

\\

1635$-$035 & 1991.4 \\
1741$-$038 & 1990.1 \\
1821$+$107\tablenotemark{a} 
           & 1991.0 \\
           & 1993.6 \\
2047$+$098 & 1993.0 \\
2319$+$272 & 1992.5 \\

\enddata

\tablenotetext{a}{Multiple potential ESEs were identified for this source.}

\end{deluxetable}

The light curve of \objectname[]{1741$-$038} shows considerable
variation both before and after the ESE identified by
\citeauthor{fjwg92}~(\citeyear{fjwg92}; cf.\ Figure~\ref{fig:ltcurv}).
This source has been identified as showing considerable refractive
scintillation \citep{hn86,dennisonetal87}.  Its light curve may be an
indication that ESEs should indeed be considered not as distinct
phenomenon, but as ``extreme refractive events'' \citep{r90} forming
the extreme cases of variability due to refractive scintillation.

We consider it likely that future monitoring programs will need to
continue relying on both systematic and visual identification of ESEs
to separate instrumental and intrinsic variations of the sources.

\section{Conclusions}\label{sec:conclude}

We have presented multi-year radio light curves for a sample of 149
sources.  The sources were monitored at two frequencies (approximately
2.5 and~8~GHz) for intervals ranging from~3 to~15~yr, with a
monitoring frequency of roughly one flux density measurement every
2~days.

We have used a variety of techniques to assess the variability of
these light curves.  The rms flux densities of the light curves are
highly correlated between the two frequency bands.  The variability is
highly correlated with spectral index, with flat-spectrum sources more
likely to be variable than steep-spectrum sources, and is more
pronounced at the higher frequency, with a frequency scaling
suggesting that the electron energy distribution in the flares has a
power-law index of~2.5.  We have also used low-order autoregressive,
integrated, moving average models (ARIMA models) to characterize the
light curves.  We find that most sources can be described by a limited
number of models.  Further, we find that the same set of models
describe sources at both frequencies.  The combination of these two
analyses suggests, though not entirely unambiguously, that these
sources share a common, broadband emission mechanism.

A structure function analysis indicates the sources display a
short-term ($\sim 10$~d) variability due to radio-wave scattering.  An
earlier analysis on a subset of these sources found a logarithmic
structure function slope of~1--1.5.  We find a mean slope of~0.3 and a
strong correlation of a source's structure function slope with its
Galactic latitude, indicative of at least a fraction of our sources
exhibiting radio-wave scattering through an extended medium.  The
lower mean slope in our larger sample as compared to that of
\citeauthor{fiedleretal87} may be indicative of the larger number of
source at high latitudes for which scattering is likely to be less
important.

The primary motivation for this monitoring program was the
identification of extreme scattering events.  In an effort to identify
ESEs in a systematic manner, we have used a wavelet transform of the
light curves to indicate intervals of significant variability.
Distinguishing instrumentally-induced variations from ESEs is
difficult for these data.  Of 7 ESEs identified previously, only 2
were found to be significant in our wavelet analysis.  Most of the
remaining 5 ESEs produced a potentially large wavelet coefficient, but 
were at epochs when other sources did as well.  Our wavelet analysis
also found 15 events in the light curves of~12 sources that we
consider to be probable ESEs.  However, distinguishing ESEs in future
monitoring programs will probably continue to require a combined
approach of both systematic and visual inspections of light curves.

We believe that this sample of 149 sources, with light curves ranging
from 6 to~15~yrs in length, presents a unique resource for studying
both intrinsic and propagation-induced variability.  The data
presented here are available on the World Wide Web at the
\anchor{http://ese.nrl.navy.mil/}{NRL GBI site},
\url{http://ese.nrl.navy.mil/}.

\acknowledgements We acknowledge the outstanding contribution of the
staff of the Green Bank Interferometer in maintaining an aging, yet
sensitive instrument.  We thank P.~Crane, A.~Fey, and B.~Rickett for
helpful discussions and J.~Imamura for assistance with
Table~\ref{tab:gbisources}.  We thank the referee for helpful
suggestions to improve the presentation of our results.  This research
made use of SIMBAD, at the CDS, Strasbourg, France, and the
Astronomical Data System.  A portion of this work was performed while
TJWL held a National Research Council-NRL Research Associateship.  The
GBI is a facility of the National Science Foundation and was operated
by the National Radio Astronomy Observatory under contract to the USNO
and NRL during these observations.  Basic research in radio astronomy
at the NRL is supported by the Office of Naval Research.

\clearpage

\begin{figure}[tbh]
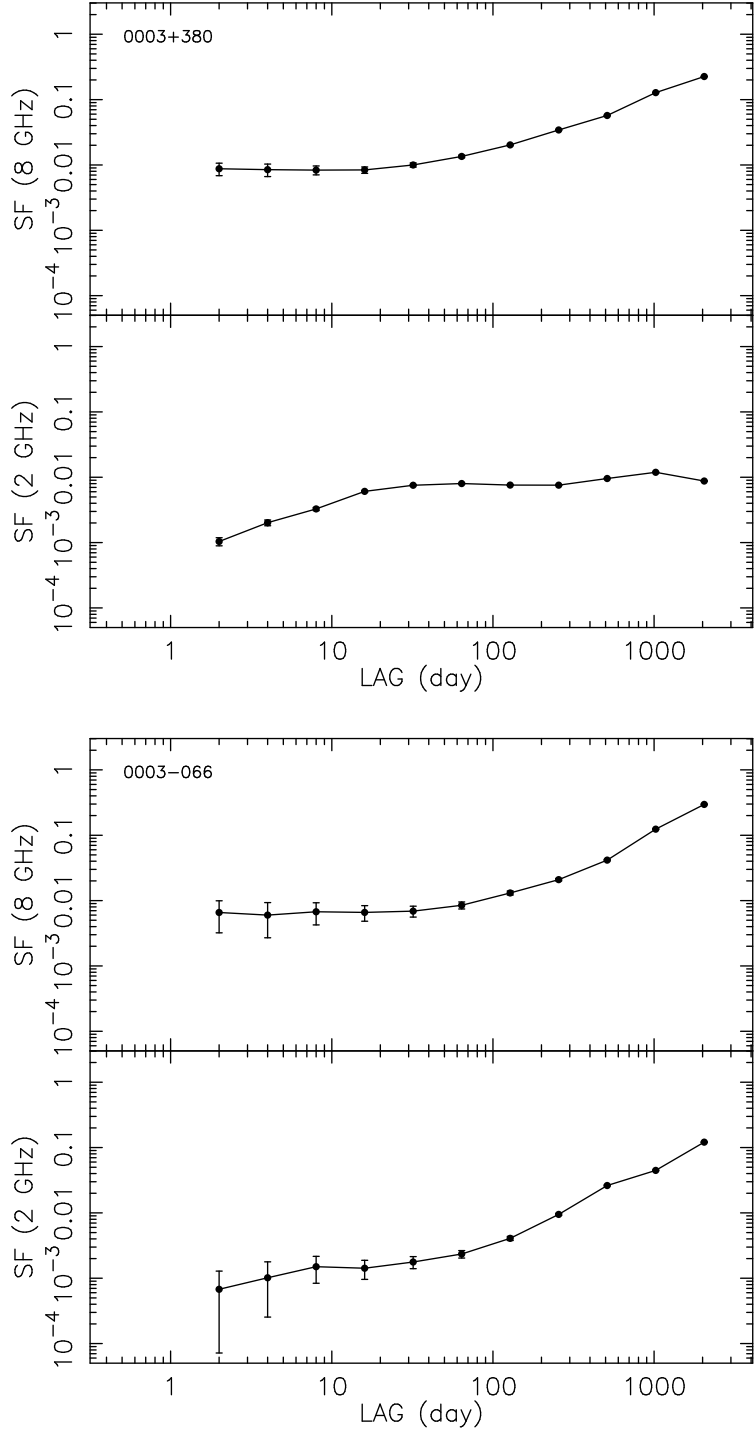

 \begin{center}
 \mbox{\psfig{file=Lazio_f9aa.ps,width=0.6\textwidth,angle=-90,silent=}}
 \linebreak\vspace{0.5ex}\linebreak
 \mbox{\psfig{file=Lazio_f9ab.ps,width=0.6\textwidth,angle=-90,silent=}}
\end{center}
\vspace{-0.5cm}
\caption[]{Structure functions.  Structure functions were computed
using equation~(\ref{eqn:sf}), using \emph{unsmoothed} light curves,
and were normalized by the variance of the light curve.  Pairs of data
on opposite sides of the change in receivers (1989~August) do not
contribute to these structure functions.  \emph{Only a portion of 
Figure~\ref{fig:sf} is shown.  All of Figure~\ref{fig:sf} and the data
used to construct it are available at
\url{http://ese.nrl.navy.mil/GBI/GBI.html}.}}
\label{fig:sf}
\end{figure}

\end{document}